# Two-dimensional Semiconductor Computational Carrier Mobility Genome

Chenmu Zhang, Yuanyue Liu*

*Texas Materials Institute and Department of Mechanical Engineering,*

*The University of Texas at Austin, Austin, Texas 78712, USA*

*Yuanyue.liu@austin.utexas.edu*

**Abstract:**

Two-dimensional (2D) crystalline semiconductors hold promise for next-generation electronic devices due to its atomical thickness and consequent properties. Despite years of search, literature-reported 2D semiconductors commonly suffered from low room-temperature charge mobility (< 200 $cm^2V^{-1}s^{-1}$), due to the dimensionality-increased "density of scattering", undesirable defects during fabrication and/or strong electron-phonon scattering. Therefore, understanding charge scatterings in 2D semiconductors via computational tools and discovering new 2D semiconductors with high mobility (> 1000 $cm^2V^{-1}s^{-1}$) are both desirable. Here we review the accurate *ab initio* approaches for electron-phonon/defect/boundary scattering developed these years, and the efforts made in high-mobility 2D semiconductor high throughput screening. Starting from these studies, the common genome of high-mobility 2D semiconductor are summarized and discussed, which would contribute to further discovering of high mobility in 2D semiconductors.

## 1. Introduction:

Two-dimensional (2D) crystalline semiconductors are semiconducting materials with a thickness of only one or few atomic layers. The extreme thinness of 2D semiconductors introduces many unique properties compared to 3D counterparts, such as high electrostatic control[1,2], optical transparency[3], and mechanical flexibility[4,5]. These properties earn 2D semiconductors immense interest for various applications in next-generation electronic devices, including optoelectronics[6], flexible nanoelectronics[4], spintronics[7], nonvolatile memory[8] and especially, 2D field-effect transistor[9-11] with small dimensions.

Finding suitable 2D semiconductors to substitute 3D semiconductor as channel material in transistor is seen to be a critical step for maintaining Moore's low. Since the 1960s, the number of transistors in a typical microprocess has followed a remarkable exponential growth, facilitated by continuously down-scaling of physical dimensions of transistors. However, when scaled down to nanoscale dimensions, the conventional 3D-semiconductor-based channels suffer from substantial performance degradation and heat dissipation due to surface roughness and dangling bonds. Specifically, the limiting carrier mobility of 3D semiconductor decreases with body thickness to the sixth power, $\mu \sim t^6$, due to strong electron scattering[12,13]. The atomically-thin 2D semiconductor without surface dangling bonds, would have intrinsic carrier mobility with little variation with thickness, and thus shed lights on further promising channel materials in scaled-down transistors[10,11].

However, in spite of its merits, the 2D semiconductor commonly suffered from relatively lower carrier mobility at room temperature. For example, $MoS_2$, one of the most common 2D semiconductor, has an intrinsic electron mobility < 200 $cm^2V^{-1}s^{-1}$, much lower than electron mobility of bulk silicon (1400 $cm^2V^{-1}s^{-1}$). Although it is often in debate whether carriers at small channel length is dominated by ballistic transport (band structure matters) or diffusion transport (mobility matters), the simulation of 2D transistor with 7-nm gate[11] indicates that a larger carrier mobility $\mu$ always leads to a larger drain current and thus a better transistor performance, and the optimal performance can be achieved by $\mu >$ 1000 $cm^2V^{-1}s^{-1}$. Therefore, the $\mu >$ 1000 $cm^2V^{-1}s^{-1}$ can be considered as a long-term target for high mobilities in 2D transistor[10]. Although very recently the record of highest room temperature mobility of monolayer 2D semiconductor (655 $cm^2V^{-1}s^{-1}$) is achieved[14] in hole-doped $WSe_2$, it is still lower than target mobility. As reviewed in Figure 4, the experimentally reported mobilities in 2D semiconductor are generally < 200 $cm^2V^{-1}s^{-1}$, far below 3D semiconductor counterparts and the target mobility. The further analysis based on first-principles calculations shows that the generally lower mobility in 2D can be attributed to the dimensionality-increased "density of scattering" (see Section 5 for details), fabrication-induced disorder scattering and/or strong electron-phonon coupling.

Indeed, the in-depth understanding of 2D semiconductor mobility is enabled by the state-of-the-art first-principles approaches developed these years. The state-resolved transition probability of electrons due to phonon, defect and boundary scattering are benefited from the development of algorithms and computational power. In conjunction with the Boltzmann transport equation, the first-principles carrier mobility of semiconductor can be obtained and shows tremendous improvement of accuracy over conventional simplified models (e.g. deformation potential theory). In addition to understanding, an accurate and efficient computational approach also helps to discover new high-mobility 2D semiconductors. Recent high-throughput calculations[15,16] show that the high intrinsic mobility (i.e. phonon-limited mobility > 1000 $cm^2V^{-1}s^{-1}$) exists in many potential 2D semiconductors. The common features of these high-mobility 2D semiconductors include small effective mass, anisotropic band edge, small Born effective charges, high LO phonon frequency, large "carrier-lattice distance" and/or weak electron-phonon couplings, which thus can be regarded as the high-mobility genomes of 2D semiconductors. The combinations of these genomes allow further discoveries of high-mobility 2D semiconductors.

The article is structured as follows. In Section 2, the iterative and approximate solutions of Boltzmann transport equation, which connects the macroscopic carrier mobility with microscopic electron occupations and scattering matrices, are reviewed. In Section 3, the first-principles calculation approaches for electron-phonon/defect/boundary scattering matrix elements are reviewed, especially for 2D semiconductors. In Section 4, the electron-phonon scattering model and featured scattering rate are discussed for better understanding of limiting factors of 2D semiconductor mobility. In Section 5, both experimentally-reported and computationally-reported carrier mobility in 2D semiconductors are reviewed. As we can see, the generally lower mobility in 2D semiconductors can be largely attributed to dimensional effect, from the view of "density of scattering". Section 6 reviews the recent works on high-throughput 2D semiconductors mobility screening, from which 14 2D semiconductor with mobility higher than bulk silicon (> 1400 $cm^2V^{-1}s^{-1}$) are suggested. In addition, the common feature and genome of high mobilities in 2D semiconductor are summarized, which would be helpful for further high-mobility discovery. In section 7, we examined the common deficiency of current computational approaches, which might account for the disagreement between experiments and computations. Indeed, the development of first-principles carrier mobility is still in progress, aiming to simulating more effects in realistic devices, including free-carrier screening, environmental dielectrics, ionized defects and etc..

## 2. Boltzmann transport equation: connecting microscopy and macroscopy

In this section and section 3, we review the first-principles calculation methods for electron-phonon (e-ph), electron-defect (e-d) and electron-boundary (e-b) scattering, especially on special techniques to deal with 2D semiconductors. The computational scheme can be generally divided into two parts: obtaining the state-resolved transition probability due to various scatterings from first-principles calculations, and then solving the diffusive Boltzmann transport equation (BTE) which connects the microscopic transition probability with macroscopic transport properties (e.g. carrier mobility, drift velocity). The complete workflow is shown in Figure 1.

### 2.1 Carrier drift mobility from Boltzmann transport equation

When a conductor/semiconductor is applied by an external electric field, the charge carriers in it will drift with an average velocity $v_\text{d}$ as a response to the field. At weak electric field, the $v_\text{d}$ of carrier grows linearly with the strength of electric field, the slope of which defines the carrier drift mobility $\mu$ (See Figure 2a). Therefore, the $\mu$ characterizes how fast charge in material can move with unit strength of electric field. There are various first-principles approaches to calculate the $\mu$. Stochastic methods like Monte Carlo simulation gives drift velocity vs electric field curve, from which the $\mu$ can be directly obtained by fitting the linear region[17-19]. The more common approach relies on solving Boltzmann transport equation (BTE) to obtain the space/state-resolved electronic occupation function $f(\mathbf{r},\mathbf{k})$. The carrier mobility is:

$$\mu = \frac{1}{V}\int d\mathbf{r}\mu(\mathbf{r}), \tag{1}$$

where $V$ is the volume of system, $\mu(\mathbf{r})$ is the space-resolved carrier mobility determined by $f(\mathbf{r},\mathbf{k})$:

$$\mu_{\alpha\beta}(\mathbf{r}) = \frac{q}{n_c\Omega_\text{uc}}\sum_n\int_\text{BZ}\frac{d\mathbf{k}}{\Omega_\text{BZ}}v_{n\mathbf{k},\alpha}\partial_\beta f_n(\mathbf{r},\mathbf{k}), \tag{2}$$

where $\alpha$ and $\beta$ are direction indices, $q$ is the charge of the carrier, $\Omega_\text{uc}(\Omega_\text{BZ})$ is the area of unit cell (Brillouin zone; BZ); $v_{n\mathbf{k}}$ is the group velocity for the electronic state with band index $n$ and wavevector $\mathbf{k}$; $n_\text{c}$ is the carrier density which is related with Fermi distribution $f^0$ and the electronic band structure through:

$$n_e = \sum_n\int_\text{BZ}\frac{d\mathbf{k}}{\Omega_\text{BZ}}f_n^0(\mathbf{k}), \quad n_e = \sum_n\int_\text{BZ}\frac{d\mathbf{k}}{\Omega_\text{BZ}}(1-f_n^0(\mathbf{k})), \tag{3}$$

where $n_e$ and $n_h$ are the concentrations for electrons and holes respectively; and $\partial_\beta f_n(\mathbf{r},\mathbf{k})$ in Eq. 2 is the linear response of occupation function $f$ for an electron at $\mathbf{r}$ in real-space and $\mathbf{k}$ in state-space under electric filed $\mathbf{E}$ along $\beta$. So in first order:

$$f_n(\mathbf{r},\mathbf{k}) = f_n^0(\mathbf{k}) + \sum_\beta \partial_\beta f_n(\mathbf{r},\mathbf{k}) E_\beta, \tag{4}$$

where $f^0$ is the equilibrium electronic Fermi distribution, $E$ is the electric field. The $f(\mathbf{r},\mathbf{k})$ satisfies the steady state BTE:

$$\mathbf{v}_{n\mathbf{k}} \cdot \nabla_\mathbf{r} f_n(\mathbf{r},\mathbf{k}) + \frac{q\mathbf{E}}{\hbar} \cdot \nabla_\mathbf{k} f_n(\mathbf{r},\mathbf{k}) = \left.\frac{\partial f_n(\mathbf{r},\mathbf{k})}{\partial t}\right|_{\text{coll}}. \tag{5}$$

The last term in Eq. 5 is the collision term due to the e-ph/e-d scattering, which can be written as:

$$\left.\frac{\partial f_n(\mathbf{r},\mathbf{k})}{\partial t}\right|_{\text{coll}} = -\sum_{\mathbf{k}'} T_{n\mathbf{k},n'\mathbf{k}'} f_n(\mathbf{r},\mathbf{k})[1 - f_{n'}(\mathbf{r},\mathbf{k}')] - T_{n'\mathbf{k}',n\mathbf{k}} f_{n'}(\mathbf{r},\mathbf{k}')[1 - f_n(\mathbf{r},\mathbf{k})]. \tag{6}$$

The $T$ is the transition probability of electronic state from $n\mathbf{k}$ to $n'\mathbf{k}'$ in an e-ph/e-d scattering event. Here we focus on e-ph scattering first. The relevant equations for e-d scattering can be found in Eq. 11. The $T$ for e-ph scattering is:

$$T_{n\mathbf{k},n'\mathbf{k}'} = \frac{2\pi}{\hbar}\frac{1}{N_{\mathbf{k}'}} \sum_\nu |g_{nn'\nu}(\mathbf{k},\mathbf{q})|^2 \left[ n_{\nu\mathbf{q}} \delta(\varepsilon_{n'\mathbf{k}'} - \varepsilon_{n\mathbf{k}} - \hbar\omega_{\nu\mathbf{q}}) + (n_{\nu\mathbf{q}} + 1)\delta(\varepsilon_{n'\mathbf{k}'} - \varepsilon_{n\mathbf{k}} + \hbar\omega_{\nu\mathbf{q}}) \right], \tag{7}$$

where $N_\mathbf{k}$ is the number of $\mathbf{k}$ grid used in the BZ sampling, $\mathbf{q}=\mathbf{k}'-\mathbf{k}$ is the wavevector for involved phonon, $v$ is its mode index, $n$ is the Bose distribution of the phonon and $g$ is the electron-phonon coupling (EPC) matrix element (see Section 3.1 for details). Combining Eqs. 4-7, the steady state BTE of $\partial f(\mathbf{r},\mathbf{k})$ can be obtained by taking derivatives of the electric field for Eq. 5:

$$[1 + \tau_{n\mathbf{k}}^0 \mathbf{v}_{n\mathbf{k}} \cdot \Delta_\mathbf{r}] \partial_\alpha f_n(\mathbf{r},\mathbf{k}) = -\frac{q}{\hbar} \tau_{n\mathbf{k}}^0 v_{n\mathbf{k},\alpha} \frac{\partial f_n^0(\mathbf{k})}{\partial \varepsilon_{n\mathbf{k}}} + \tau_{n\mathbf{k}}^0 \sum_{\mathbf{k}'} \tilde{T}_{n\mathbf{k},n'\mathbf{k}'} \partial_\alpha f_n(\mathbf{r},\mathbf{k}'), \tag{8}$$

where $\tau_{n\mathbf{k}}^0$ is self-energy relaxation time for electronic state $n\mathbf{k}$: $1/\tau_{n\mathbf{k}}^0 = \sum_{n'\mathbf{k}'} \bar{T}_{n\mathbf{k},n'\mathbf{k}'}$ and $\bar{T}, \tilde{T}$ for e-ph scattering are:

$$\bar{T}_{n\mathbf{k},n'\mathbf{k}'} = \frac{1 - f_{n'}^0(\mathbf{k}')}{1 - f_n^0(\mathbf{k})} T_{n\mathbf{k},n'\mathbf{k}'}, \qquad \tilde{T}_{n\mathbf{k},n'\mathbf{k}'} = \frac{f_n^0(\mathbf{k})}{f_{n'}^0(\mathbf{k}')} T_{n\mathbf{k},n'\mathbf{k}'}. \tag{9}$$

By inserting the explicit form of equilibrium distribution of electrons ($f^0$) and phonons ($n$), the $\bar{T}, \tilde{T}$ can be written in more used forms:

$$\begin{aligned}
\bar{T}_{n\mathbf{k},n'\mathbf{k}'} &= \frac{2\pi}{\hbar}\frac{1}{N_{\mathbf{k}'}} \sum_\nu |g_{nn'\nu}(\mathbf{k},\mathbf{q})|^2 [(n_{\nu\mathbf{q}} + f_{n'}^0(\mathbf{k}'))\delta(\varepsilon_{n\mathbf{k}} + \hbar\omega_{\nu\mathbf{q}} - \varepsilon_{n'\mathbf{k}'}) \\
&\quad + (1 + n_{\nu\mathbf{q}} - f_{n'}^0(\mathbf{k}'))\delta(\varepsilon_{n\mathbf{k}} - \hbar\omega_{\nu\mathbf{q}} - \varepsilon_{n'\mathbf{k}'})], \\
\tilde{T}_{n\mathbf{k},n'\mathbf{k}'} &= \frac{2\pi}{\hbar}\frac{1}{N_{\mathbf{k}'}} \sum_\nu |g_{nn'\nu}(\mathbf{k},\mathbf{q})|^2 [(1 + n_{\nu\mathbf{q}} - f_{n'}^0(\mathbf{k}'))\delta(\varepsilon_{n\mathbf{k}} + \hbar\omega_{\nu\mathbf{q}} - \varepsilon_{n'\mathbf{k}'}) \\
&\quad + (n_{\nu\mathbf{q}} + f_{n'}^0(\mathbf{k}'))\delta(\varepsilon_{n\mathbf{k}} - \hbar\omega_{\nu\mathbf{q}} - \varepsilon_{n'\mathbf{k}'})].
\end{aligned} \tag{10}$$

For e-d scattering, the $T, \bar{T}, \tilde{T}$ are much simpler as e-d scattering does not change the electronic state energy:

$$T_{n\mathbf{k},n'\mathbf{k}'} = \bar{T}_{n\mathbf{k},n'\mathbf{k}'} = \tilde{T}_{n\mathbf{k},n'\mathbf{k}'} = \frac{2\pi}{\hbar}\frac{n_{\mathrm{at}}C_{\mathrm{d}}}{N_{\mathbf{k}'}}|M_{nn'}(\mathbf{k},\mathbf{k}')|^2\,\delta(\varepsilon_{n'\mathbf{k}'}-\varepsilon_{n\mathbf{k}}), \tag{11}$$

where $n_{\mathrm{at}}$ is the number of atoms in the unit cell, $C_{\mathrm{d}}$ is the defect concentration and $M$ is the electron-defect-interaction (EDI) matrix element which will be discussed in Section 3.2.

The linearized BTE for carriers in steady state in Eq. 8 is the key equation for solving the carrier perturbation distribution $\partial f(\mathbf{r},\mathbf{k})$. With $\partial f(\mathbf{r},\mathbf{k})$ and Eq. 4, the carrier occupation function $f(\mathbf{r},\mathbf{k})$ can be obtained which will give macroscopic transport properties we concern. However, solving $\partial f(\mathbf{r},\mathbf{k})$ in Eq. 8 requires intensive efforts as it has both spatial and momentum degrees of freedom. Here we focus on two kinds of approaches for solving BTE in Eq. 8: one is iteration method which requires the information of matrix element ($g$ or $M$) and one with higher hierarchy only requires state-resolved relaxation time (e.g. $\tau^0$ in Eq. 8) at the expanse of more assumptions and lower accuracy.

## 2.2. Iterative solution for BTE

The steady-state BTE in Eq. 8 properly separates the spatial and state dependence into the left-hand and the right-hand side. Therefore, it is convenient to solve the BTE in spatial and momentum space recursively and iteratively[20,21] via Eq. 8. The $\partial f(\mathbf{r},\mathbf{k})$ can be expressed as a summation:

$$\partial_\alpha f_n(\mathbf{r},\mathbf{k}) = \partial_\alpha f_n^0(\mathbf{r},\mathbf{k}) + \partial_\alpha f_n^1(\mathbf{r},\mathbf{k}) + \partial_\alpha f_n^2(\mathbf{r},\mathbf{k}) + \ldots, \tag{12}$$

where different order of $\partial f(\mathbf{r},\mathbf{k})$ can be obtained by solving:

$$\begin{aligned}
&[1+\tau_{n\mathbf{k}}^0 \mathbf{v}_{n\mathbf{k}}\cdot\Delta_{\mathbf{r}}]\partial_\alpha f_n^0(\mathbf{r},\mathbf{k}) = -\frac{q}{\hbar}\tau_{n\mathbf{k}}^0 v_{n\mathbf{k},\alpha}\frac{\partial f_n^0(\mathbf{k})}{\partial\varepsilon_{n\mathbf{k}}},\\
&[1+\tau_{n\mathbf{k}}^0 \mathbf{v}_{n\mathbf{k}}\cdot\Delta_{\mathbf{r}}]\partial_\alpha f_n^i(\mathbf{r},\mathbf{k}) = \tau_{n\mathbf{k}}^0\sum_{n'\mathbf{k}'}\tilde{T}_{n\mathbf{k},n'\mathbf{k}'}\partial_\alpha f_n^{i-1}(\mathbf{r},\mathbf{k}).
\end{aligned} \tag{13}$$

By inserting Eqs. 12 and 13 into Eq. 8, it can be proved that the $\partial f(\mathbf{r},\mathbf{k})$ from Eq. 12 satisfies the linearized BTE in Eq. 8 and consequently, the corresponding $f(\mathbf{r},\mathbf{k})$ (Eq. 4) is indeed the solution of BTE. For an infinite system without boundaries, the electron occupation function $f(\mathbf{k})$ is independent of spatial coordinates and thus the linear response is $\partial f(\mathbf{k})$ can be solved by:

$$\begin{aligned}
&\partial_\alpha f_n^0(\mathbf{k}) = -\frac{q}{\hbar}\tau_{n\mathbf{k}}^0 v_{n\mathbf{k},\alpha}\frac{\partial f_n^0(\mathbf{k})}{\partial\varepsilon_{n\mathbf{k}}},\\
&\partial_\alpha f_n^i(\mathbf{k}) = \tau_{n\mathbf{k}}^0\sum_{n'\mathbf{k}'}\tilde{T}_{n\mathbf{k},n'\mathbf{k}'}\partial_\alpha f_{n'}^{i-1}(\mathbf{k}').
\end{aligned} \tag{14}$$

The solution of $\partial f(\mathbf{k})$ in Eq. 14 involves repeated multiplication of scattering matrix $\tilde{T}$ and the state-resolved $\partial f(\mathbf{k})$, which is equivalent to solution in Ref. 22. When spatial dependence is considered, the iteration in Eq. 13 requires solving a differential equation in real space:

$$\left[1+\tau_0 v_z\frac{\partial}{\partial z}\right]\partial f^i(z) = g^{i-1}(z), \tag{15}$$

where $z$ is selected as the direction of $\mathbf{v}_{n\mathbf{k}}$, $g^{i-1}(z)$ represents the functions on the right sides of Eq. 13 and the state index $n\mathbf{k}$ is omitted for simplicity. The $\partial f^i$ in Eq. 15 can be solved on real-space grids:

$$\partial f^i(z) = \partial f^i(z_0)\exp(-\frac{z-z_0}{v_z\tau_0}) + \int_{z_0}^{z} dz' \frac{g^{i-1}(z')}{v_z\tau_0}\exp(-\frac{z-z'}{v_z\tau_0}). \tag{16}$$

By using the $\tilde{T}$ from Eq. 10 (e-ph) or from Eq. 11 (e-d), the combination of Eq. 12, 13, 16 gives the final solution of $\partial f(\mathbf{r},\mathbf{k})$ due to e-ph or e-d scattering and then the corresponding macroscopic carrier mobility can be obtained from Eqs. 1 and 2. The e-b scattering affects the mobility by posing boundary effect on $\partial f(\mathbf{r},\mathbf{k})$ at boundaries:

$$\partial_\alpha f_n^i(\mathbf{r}=\mathbf{r}_\mathrm{b},\mathbf{k}') = \sum_{n\mathbf{k}} P_{n\mathbf{k},n'\mathbf{k}'} \partial_\alpha f_n^i(\mathbf{r}=\mathbf{r}_\mathrm{b},\mathbf{k}), \tag{17}$$

where $\mathbf{r}_\mathrm{b}$ is the position of boundaries in system, $P$ is the transition probability for an electron scattered from initial state $n\mathbf{k}$ to final state $n'\mathbf{k}'$ due to boundaries, which would be discussed in Section 3.3.

## 2.3. Relaxation time approximation in BTE

Although the iteration approach solves the BTE accurately, it requires the storage and multiplication of large state-to-state transition matrix (see $\tilde{T}$ in Eqs. 13 and 14), which is usually computationally intensive. An BTE solution in balance of efficiency and accuracy is desirable. In practice, two relaxation time approximations (RTA) for solving the steady state BTE, which do not require the knowledge of $\tilde{T}$ or iteration, are widely used. Here we illustrate how to apply these two RTAs in infinite system or finite film with parallel boundaries (or ribbon for 2D systems). In finite films and ribbons, the electron occupation function is independent of spatial coordinates parallel to boundaries due to the translation symmetry. Therefore, the $f(\mathbf{r},\mathbf{k})$ and $\partial f(\mathbf{r},\mathbf{k})$ can be written as $f(z,\mathbf{k})$ and $\partial f(z,\mathbf{k})$ where $z$ is the norm of the boundary. Within RTAs, the $\partial f(z,\mathbf{k})$ is assumed to have such format:

$$\begin{aligned}\partial_\alpha f_n(z,\mathbf{k}) &= -q\tau_{n\mathbf{k}} v_{n\mathbf{k},\alpha} \frac{\partial f_n^0(\mathbf{k})}{\partial \varepsilon_{n\mathbf{k}}} Z_n(z,\mathbf{k}) \\ &= -q\tau_{n\mathbf{k}} v_{n\mathbf{k},\alpha} \frac{\partial f_n^0(\mathbf{k})}{\partial \varepsilon_{n\mathbf{k}}} \left[1 + F_n(\mathbf{k})\exp\left(-\frac{z}{\tau_{n\mathbf{k}} v_{n\mathbf{k},z}}\right)\right],\end{aligned} \tag{18}$$

where $F_n(\mathbf{k})$ is state-resolved function to be determined by e-b scattering boundary conditions[23] in Eq. 17, $\tau$ is an unknown relaxation time specific to electronic state. The specific format of $\tau$ and the solidity of relaxation time assumption in Eq. 18 can be verified by inserting Eq. 18 back to BTE in Eq. 8. With the help of identity:

$$\frac{\partial f_n^0(\mathbf{k})}{\partial \varepsilon_{n\mathbf{k}}} = -\frac{f_n^0(\mathbf{k})(1-f_n^0(\mathbf{k}))}{k_\mathrm{B}T}, \tag{19}$$

and the detailed balance condition:

$$\bar{T}_{n'\mathbf{k}',n\mathbf{k}} f^0_{n'}(\mathbf{k}')[1-f^0_n(\mathbf{k})] = \bar{T}_{n\mathbf{k},n'\mathbf{k}'} f^0(\mathbf{k})[1-f^0_{n'}(\mathbf{k}')], \tag{20}$$

another format of BTE from Eq. 4, 5 and 6 can be obtained under the RTAs in Eq. 18:

$$\frac{1}{\tau_{n\mathbf{k}}} = \sum_{n'\mathbf{k}'} \bar{T}_{n\mathbf{k},n'\mathbf{k}'} \left[ 1 - \frac{\mathbf{v}_{n\mathbf{k}} \cdot \mathbf{v}_{n'\mathbf{k}'}}{|\mathbf{v}_{n\mathbf{k}}|^2} \frac{\tau_{n'\mathbf{k}'}}{\tau_{n\mathbf{k}}} \frac{Z_{n'}(z,\mathbf{k}')}{Z_n(z,\mathbf{k})} \right]. \tag{21}$$

Two approximations can be made based on the different processing of the square bracket in Eq. 21. By assuming the whole term in square bracket as 1, then we have $1/\tau_{n\mathbf{k}} = \sum_{n'\mathbf{k}'} \bar{T}_{n\mathbf{k},n'\mathbf{k}'}$ =and consequently the $\tau$ is equal to self-energy relaxation time $\tau^0$, which is call self-energy relaxation time approximation (SERTA). A better choice is to assume that $\tau_{n\mathbf{k}} Z_n(z,\mathbf{k}) \approx \tau_{n'\mathbf{k}'} Z_{n'}(z,\mathbf{k}')$ and to preserve the angle term related to **v**:

$$\frac{1}{\tau^m_{n\mathbf{k}}} = \sum_{n'\mathbf{k}'} \bar{T}_{n\mathbf{k},n'\mathbf{k}'} \left[ 1 - \frac{\mathbf{v}_{n\mathbf{k}} \cdot \mathbf{v}_{n'\mathbf{k}'}}{|\mathbf{v}_{n\mathbf{k}}||\mathbf{v}_{n'\mathbf{k}'}|} \right] = \sum_{n'\mathbf{k}'} \bar{T}_{n\mathbf{k},n'\mathbf{k}'} (1-\cos\theta_{n\mathbf{k},n'\mathbf{k}'}), \tag{22}$$

where a more symmetric angle term, $(1-\cos\theta_{n\mathbf{k},n'\mathbf{k}'})$, is used to ensure the $\tau^m$ is always positive. The scattering rate in Eq. 22 reduces the scattering contribution from forward-scattering (by 1-cos$\theta$=0, which does not change the momentum) and increases the weight of back-scattering (by 1-cos$\theta$=2, which reverses the momentum), thus characterizing the scattering rate of momentum in materials. Therefore, the Eq. 22 is called momentum relaxation time approximation (MRTA) and $\tau^m$ is momentum relaxation time, which usually give more accurate results than SERTA, especially for materials with strong small-angle scattering. In a study over 54 different 3D semiconductors, the mean-absolute-percentage-error of MRTA mobility is 18% and is 48% for SERTA, when compared to more accurate iterative mobility[24]. In 2D semiconductors, the mobility difference from MRTA and iteration methods is 0.5% in $MoS_2$ and 5% in InSe monolayer, smaller than that from SERTA (12% in $MoS_2$ and 50% in InSe)[25]. The MRTA also gives close results to iteration method when spatial-dependence in $f(z,\mathbf{k})$ is considered. In copper film with (111) and (110) surfaces, the film resistivity difference between MRTA and iterative method is less than 3% when film thickness ranges from 4 nm to 400 nm. Therefore, the MRTA can be viewed as a reliable and efficient solution to BTE, which gives the mobility of film as:

$$\mu^m_{\alpha\beta} = \frac{1}{A}\int_0^A dz\, \mu^m_{\alpha\beta}(z) = \frac{1}{A}\int_0^A dz \frac{q}{n_c \Omega_{\text{uc}}} \sum_n \int_{\text{BZ}} \frac{d\mathbf{k}}{\Omega_{\text{BZ}}} \tau^m_{n\mathbf{k}} v_{n\mathbf{k},\alpha} v_{n\mathbf{k},\beta} \frac{\partial f^0_n(\mathbf{k})}{\partial \varepsilon_{n\mathbf{k}}} \left[ 1 + F_n(\mathbf{k}) \exp\left( -\frac{z}{\tau^m_{n\mathbf{k}} v_{n\mathbf{k},z}} \right) \right], \tag{23}$$

where $A$ is the thickness of film, $\tau^m$ is the momentum relaxation time (Eq. 22) due to e-ph scattering (Eq. 10) or e-d scattering (Eq. 11), and $F$ is determined by boundary conditions due to e-b scattering (Eq. 17). For an infinite material without e-b scattering, the MRTA mobility $\mu^m$ becomes:

$$\mu^m_{\alpha\beta} = \frac{q}{n_c \Omega_{\text{uc}}} \sum_n \int_{\text{BZ}} \frac{d\mathbf{k}}{\Omega_{\text{BZ}}} \tau^m_{n\mathbf{k}} v_{n\mathbf{k},\alpha} v_{n\mathbf{k},\beta} \frac{\partial f^0_n(\mathbf{k})}{\partial \varepsilon_{n\mathbf{k}}}. \tag{24}$$

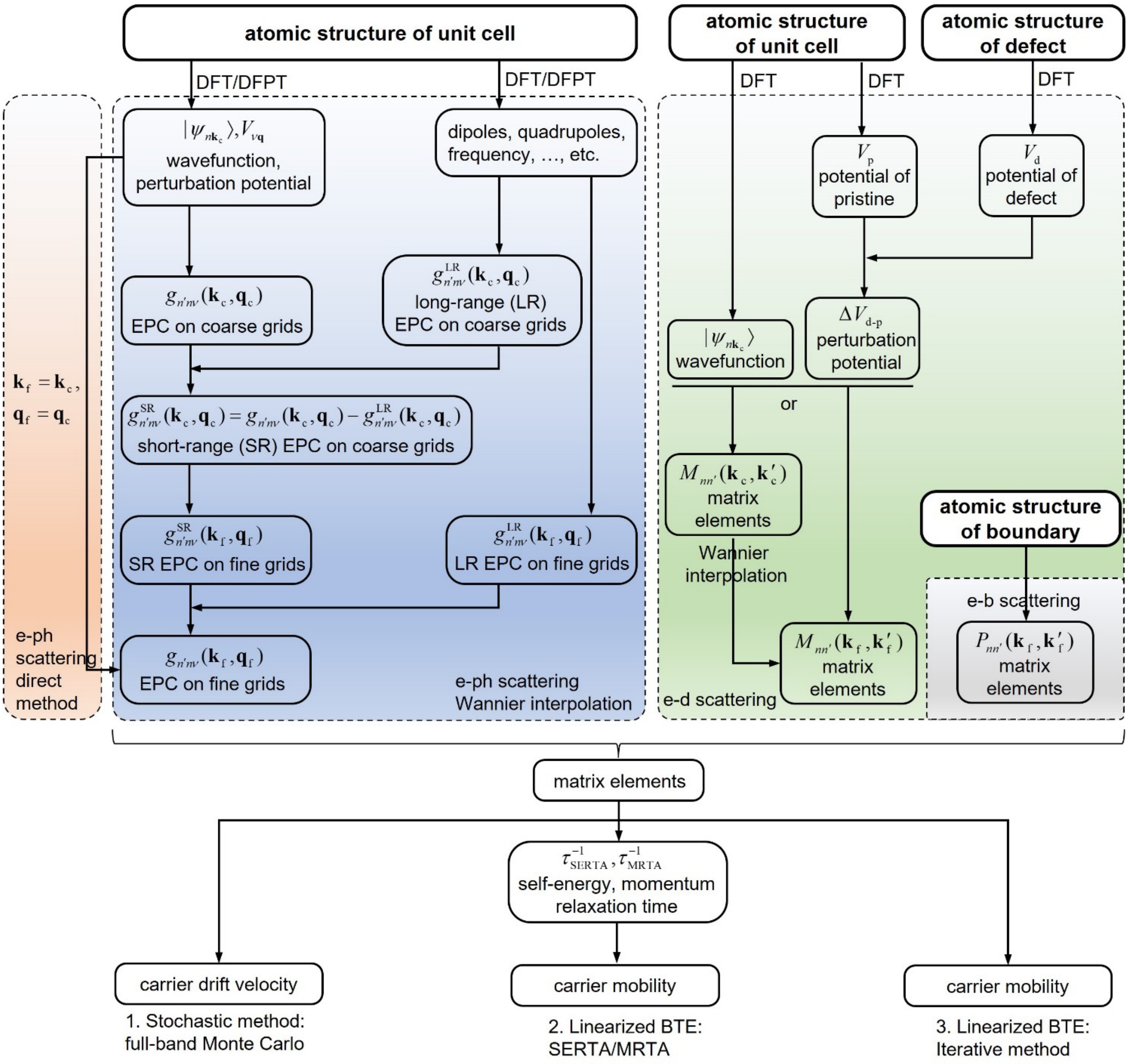

**Figure 1** Computation flowchart of first-principles carrier mobility.

## 3. First-principles calculations on transition matrix

In this section we review the first-principles approaches of calculating scattering matrix elements for e-ph (EPC $g$ matrix), e-d (EDI $M$ matrix) and e-b (transition probability $P$) scattering, especially with the focus on techniques developed for 2D semiconductor.

### 3.1. Electron-phonon scattering

The key quantities for the transition matrix $\bar{T}$ in e-ph scattering is EPC matrix element $g$ (see Eq. 10), which is defined as:

$$g_{n'n\nu}(\mathbf{k},\mathbf{q}) = \langle \psi_{n'\mathbf{k}+\mathbf{q}} | V_{\nu\mathbf{q}} | \psi_{n\mathbf{k}} \rangle, \tag{25}$$

where $\psi_{n\mathbf{k}}$, $\psi_{n'\mathbf{k}+\mathbf{q}}$ are the initial electronic state and final state respectively and $V_{\nu\mathbf{q}}$ is the perturbation potential induced by the phonon $\nu\mathbf{q}$. The $V_{\nu\mathbf{q}}$ has a more specific expression:

$$V_{\nu\mathbf{q}} = e^{i\mathbf{q}\cdot\mathbf{r}} l_{\nu\mathbf{q}} \sum_{\kappa\alpha} \sqrt{\frac{M_0}{M_\kappa}} e_{\kappa\alpha,\nu}(\mathbf{q}) \frac{\partial V_{\mathrm{KS}}^{(\mathrm{ph})}}{\partial u_{\kappa\alpha}(\mathbf{q})}, \tag{26}$$

where $M_0$ is a reference mass (usually selected as the proton mass in Refs. [26,27] but chose as the mass of unit cell in this article), $M_\kappa$ is the mass of atom $\kappa$ in unit cell, $u_{\kappa\alpha}(\mathbf{q})$ is the periodic displacement of atom $\kappa$ in $\alpha$ direction, $e_{\kappa\alpha,\nu}(\mathbf{q})$ is the corresponding eigenvector in phonon $\nu\mathbf{q}$ and $l_{\nu\mathbf{q}}$ is the zero-point vibration amplitude:

$$l_{\nu\mathbf{q}} = \sqrt{\frac{\hbar}{2M_0\omega_{\nu\mathbf{q}}}}. \tag{27}$$

The $\frac{\partial V_{\mathrm{KS}}^{(\mathrm{ph})}}{\partial u_{\kappa\alpha}(\mathbf{q})}$ =in Eq. 26 is the derivative of Kohn-Sham (KS) potential due to periodic atom displacement directly calculated from DFPT and then the $V_{\nu\mathbf{q}}$ and then $g$ can be calculated with eigenvector ($e$) and frequency ($\omega$), which usually accumulates computational errors. Therefore, the EPC strength $D$:

$$D_{n'n\nu}(\mathbf{k},\mathbf{q}) = \frac{g_{n'n\nu}(\mathbf{k},\mathbf{q})}{l_{\nu\mathbf{q}}}, \tag{28}$$

is a more reliable quantity in EPC calculations and is more used for comparison in literatures.

There are various approaches to calculate the EPC matrix element from first-principles calculations, including finite-difference method[28] (usually referred as “frozen-phonon” method or “supercell” method), dielectric method[29,30] and density functional perturbation theory (DFPT) method[31-33]. The DFPT calculation can be performed within the unit cell which is more efficient than the frozen-phonon method and the EPC calculations for phonons with different **q** are decoupled, which expedites the band-edge transport calculations where only partial **q** grids are involved. Meanwhile, the self-consistent approach based on Sternheimer equation in DFPT[31,33] enables it to be performed on occupied electronic states, which avoids the band convergence problem and reduces the computational cost. In the following, we focus on DFPT approach, especially the special techniques developed for 2D semiconductors.

The method using DFPT to directly compute all necessary EPC $g$ matrix for transport properties is called “direct method” in the article. In 3D system. the “direct method” is limited by the **q** grid sampling since individual DFPT calculation is required for each $g$ with different phonon wavevector **q**. In bulk GaAs, $10^6$ **q** points and thus the same number of individual DFPT calculations are required for converged relaxation time[34], which makes the direct method prohibited for bulk systems. However, in 2D systems with lower dimensionality, by taking full use of crystal symmetries and nonuniform **q** grid sampling, the number of DFPT calculations for carrier mobility can be reduced to ~ 200, making the direct method a competitive approach for 2D materials[35].

A necessary modification in DFPT for 2D semiconductor is the Coulomb cutoff along the nonperiodic direction. In many first-principles tools, the 2D systems are simulated in periodic boundary conditions, which leads to fictitious interlayer interaction. Especially in DFPT calculations, the longitudinal optical (LO) phonon electrizes the polar 2D semiconductors with periodic positive and negative charges and results in an electrostatic perturbation potential $V$ (and the corresponding $g$) scaling as $1/|\mathbf{q}|$ when $|\mathbf{q}|\rightarrow 0$. Note that the diverged $g$ (~ $1/|\mathbf{q}|$) cannot be avoided by merely increasing the interlayer distance in simulations, and thus the truncation of Coulomb interaction (i.e. 2D Coulomb cutoff) along the norm direction is essential for DFPT calculations for 2D semiconductor[36].

The efficiency of calculation of first-principles $g$ can be further improved by various interpolation approaches, including Fourier interpolation[37,38] of perturbation potential $V$ and Wannier interpolation[39] of EPC $g$ matrix. In semiconductors, the singularity of $V$ and $g$ at $\mathbf{q}$=0 makes the interpolation nontrivial and thus requires special treatments. The basic idea is to separate the part with singularity in $V$ and $g$, which is called long-range (LR) part and has explicit expressions based on physical models, and to use reciprocal/real-space transform (i.e. Fourier transform for $V$ and Wannier transform for $g$) to interpolate the remaining short-range (SR) part. As shown in flowchart in Fig. 1, the $g^{\mathrm{LR}}$ with singularity can be explicitly computed at fine grids while the $g^{\mathrm{SR}}$ is calculated by Wannier interpolation, the summation of which gives the full $g$ at interpolated fine **k**/**q** grids.

Therefore, the explicit expression of LR part of $V$ and $g$ is crucial for successful interpolation and accurate carrier mobility. According to the order of multiple expansion used in physical model, the LR scattering can be divided into dipolar, quadrupolar and higher order scattering. The missing of dipolar or quadrupolar scattering in interpolation would lead to several times or 20-50% mobility difference for both 3D[40] and 2D semiconductors[41]. The dipolar and quadrupolar scattering in 3D semiconductors are implemented in 2015[42] and 2020[37,38,43,44], respectively, while the quadrupolar scattering in 2D semiconductors is tackled very recently[41,45]. By appropriately separating the LR electrostatic interaction starting from Coulomb kernel[46], the $g^{\mathrm{LR}}(\mathbf{k},\mathbf{q})$ in 2D semiconductor can be approximated by:

$$g_{n'n\nu}^{\mathrm{LR}}=\frac{ie^2}{2\Omega_{\mathrm{uc}}}\sum_{\kappa}\left(\frac{\hbar}{2M_{\kappa}\omega_{\nu\mathbf{q}}}\right)^{1/2} f(|\mathbf{q}|)\frac{\mathbf{q}\cdot\mathbf{Z}_{\kappa}(\mathbf{q})\cdot\mathbf{e}_{\kappa\nu}(\mathbf{q})}{|\mathbf{q}|\epsilon^{\parallel}(\mathbf{q})}\langle\psi_{n'\mathbf{k}+\mathbf{q}}|e^{i\mathbf{q}\cdot(\mathbf{r}-\boldsymbol{\tau}_{\kappa})}|\psi_{n\mathbf{k}}\rangle, \tag{29}$$

where $\Omega_{\mathrm{uc}}$ is the unit cell care, $\kappa$ is the index of the atom in the unit cell, $M$ is the atomic mass, $\mathbf{e}_{\nu}(\mathbf{q})$ is the eigenvector of phonon $\nu\mathbf{q}$, $\boldsymbol{\tau}$ is the atomic position, $\epsilon^{\parallel}$ is the macroscopic in-plane dielectric function and $\psi_{n\mathbf{k}}$, $\psi_{n'\mathbf{k}+\mathbf{q}}$ are the initial and final electronic state respectively. The $f(q)$=1-tanh($qL$/2) is range separation function and $L$ is range separation length, which can be optimized from interatomic force constants[46]. The $\mathbf{Z}_{\kappa}(\mathbf{q})$ is a key matrix describing charge response to nuclear displacement and can be expanded around $\mathbf{q}$=0 as:

$$Z_{\kappa\alpha}^{\beta}(\mathbf{q}) = \hat{Z}_{\kappa\alpha}^{(\beta)} - i\sum_{\gamma}\frac{q_{\gamma}}{2}\left(\hat{Q}_{\kappa\alpha}^{\beta\gamma} - \delta_{\beta\gamma}\hat{Q}_{\kappa\alpha}^{zz}\right). \tag{30}$$

Here $\alpha$, $\beta$, $\gamma$ are direction indices that go through $x$, $y$ and $z$ ($x$ and $y$ are parallel to the basal plane of the 2D crystal, while $z$ is perpendicular), and $\hat{Z}$ and $\hat{Q}$ are Born charges and quadrupoles for 2D system[46]. If $\hat{Q}$ in Equation Eq. 30 is set as zero, then the $g^{\text{LR}}$ will only contain dipolar scattering. It should be noted that the electric field term, the Berry connection term and the out-of-plane EPC are neglected in Eq. 29, due to their relatively small contribution[45] to the $g^{\text{LR}}$.

In spite of approximations mentioned above, the interpolated $g$ in 2D semiconductor, combined from Wannier interpolated $g^{\text{SR}}$ and explicitly calculated $g^{\text{LR}}$ from Eq. 29, shows good consistency with direct DFPT calculations. In Figure 2, the EPC strength $D$ (recall that $D_{n'n\nu}(\mathbf{k},\mathbf{q}) = g_{n'n\nu}(\mathbf{k},\mathbf{q}) / l_{\nu\mathbf{q}}$ is more convenient for EPC comparison) interpolated with (solid lines) and without quadrupoles (dashed lines) are compared with that from direct DFPT calculations (dots) for selected phonon modes in intrinsic $MoS_2$ and InSe[25]. For both materials, the initial state is selected to be 30 meV above the CBM along the Γ-M direction, and the final states are located at the iso-energy circle of the same valley. Here impact of quadrupoles scattering is focused since it is less examined. As shown in Figure 2, the interpolated $D$ without quadrupoles (dashed lines) significantly deviates from the DFPT calculated ones (dots), especially for TA mode for $MoS_2$ (Figure 2c), LA and TA modes for InSe (Figure 2e), and $ZO_1$ and $ZO_2$ modes for InSe (Figure 2f). After incorporation of quadrupoles, the interpolated $D$ (solid lines) become more consistent with the DFPT calculated ones, indicating the importance of quadrupole scattering in various phonon modes. In addition, the room temperature electron mobility for $MoS_2$ and InSe calculated by three approaches: without dipole or quadrupoles (denoted as "wo. D/Q"), only with dipoles (denoted by "D") and with dipoles and quadrupoles ("D+Q"), are compared in Figure 2b. For $MoS_2$, the mobilities from different approaches range from 189 to 136 $cm^2V^{-1}s^{-1}$, with a variation of 50 %. While for InSe, the mobility decreases from 414 to 103 $cm^2V^{-1}s^{-1}$ after the incorporation of dipoles and then slightly increases to 117 $cm^2V^{-1}s^{-1}$ when both dipoles and quadrupoles are considered. It suggests that the dipoles are essential for accurate EPC interpolation in polar 2D semiconductor and quadrupoles also play a non-negligible role in mobility.

### 3.2. Electron-defect scattering

The EPC $g$ matrix counterpart in e-d scattering is the electron-defect interaction (EDI) matrix element $M$, which is defined as:

$$M_{nn'}(\mathbf{k},\mathbf{k}') = \langle \psi_{n'\mathbf{k}'} | \Delta V_{\text{d-p}} | \psi_{n\mathbf{k}} \rangle, \tag{31}$$

where $\psi_{n\mathbf{k}}$, $\psi_{n'\mathbf{k}'}$ are the initial electronic state and final state respectively and $\Delta V_{\text{d-p}}$ is the perturbation potential due to the defect. The $\Delta V_{\text{d-p}}$ can be calculated from first-principles calculations by:

$$\Delta V_{\text{d-p}} = V_{\text{KS}}^{(\text{d})} - V_{\text{KS}}^{(\text{p})}, \tag{32}$$

where $V_{\text{KS}}^{(\text{d})}, V_{\text{KS}}^{(\text{p})}$ are KS potential of a system containing defect and a pristine one without defect. In practice, the $V_{\text{KS}}^{(\text{p})}$ and $\psi_{n\mathbf{k}}$ can be obtained from the density functional theory (DFT) calculation on unit cell due to the periodicity of pristine crystal and the $V_{\text{KS}}^{(\text{d})}$ is calculated from a supercell. The flowchart of EDI matrix element calculation is illustrated in Figure 1.

Early studies on EDI *M* matrix used "all-supercell" method, which calculates all the necessary quantities ($V_{\text{KS}}$ and $\psi$) via supercell[47-49]. The "all-supercell" method is limited by the efficiency and thus prohibited for transport calculations involving numerous *M* ($\sim 10^8/10^4$ in 3D/2D semiconductors). In 2019, great speedup is demonstrated in bulk silicon by adopting most of calculation in unit cell[50] and the efficiency can be further enhanced by Wannier interpolation of *M* from coarse $\mathbf{k}_c$ grid to fine $\mathbf{k}_f$ grid[51] (see diagram in Fig. 1). However, for 2D semiconductors, the EDI computation for carrier transport is only available very recently [defect xx]. In Ref. [defect xx], the EDI *M* matrix due to X vacancy in $MX_2$ (denoted as $V_X$; sulfur vacancy for $MoS_2$ and $WS_2$, selenium vacancy for $MoSe_2$ and $WSe_2$) are calculated and the corresponding defect-limited mobility is compared with the phonon-limited mobility.

As we can see in Figure 2g, the EDI *M* matrix of $V_X$ in $MX_2$ generally shows a decreasing trend with increase of scattering wavevector length for intravalley scattering, with initial and final states located at the iso-energy circle of the same valley 25 meV above CBM. The K/K' intervalley e-d scattering for $V_X$ defects is prohibited in $MX_2$ due to the spin-locking and 3-fold rotation symmetry[48] and thus leading to negligible contributions to the mobility. Across different $MX_2$, the $MoSe_2$ has the largest EDI strength, followed by $WSe_2$, $WS_2$ and $MoS_2$, and accordingly, the $MoSe_2$ has the smallest momentum relaxation time ($MoSe_2 < WSe_2 < WS_2 < MoS_2$; see relaxation time in Figure 2h). In Figure 2i, the $V_X$-limited mobility ($\mu^{(\text{d})}$) of $MX_2$ are compared with the phonon-limited mobility ($\mu^{(\text{ph})}$), at a represented defect concentration $n_V = 10^{12}$ cm$^{-2}$. It is worth noting, most of $\mu^{(\text{d})}$ are comparable to $\mu^{(\text{ph})}$, indicating significant contributions from e-d scattering in $MX_2$ experimentally-reported mobilities as defect concentration usually[52,53] $> 10^{12}$ cm$^{-2}$. For $WSe_2$ and $WS_2$ holes, $\mu^{(\text{ph})}$ is 5 times of $\mu^{(\text{d})}$, which means the limiting factor of room temperature mobility is more likely to be e-d scattering instead of e-ph scattering. The study on $V_X$ limited mobility highlights the e-d scattering in 2D semiconductors and inspires further studies on this unexplored field in 2D semiconductors having various types of defect and defect concentrations.

### 3.3. Electron-boundary scattering

Although the carrier transport in 2D semiconductors is protected by its atomically flat surfaces and free of dangling bonds on surfaces, the electrons are still likely to be scattered by the lateral boundaries, including line defect, grain boundary and surface. The electron-boundary (e-b) scattering leads to observable hydrodynamic electron flow and in-plane resistivity anisotropy in 2D materials[54-57] like Graphene, $WTe_2$ and $PdCoO_2$, however, remains elusive for conducting 2D semiconductors. In addition, there is a lack of calculation approach to appropriately incorporate e-b scattering with the e-ph/e-d scattering: atomistic simulations based on Green's function or other quantum transport approaches is limited by the scaling of the number of atoms simulated and is mainly focused on ballistic transport without e-ph/e-d scattering[58-61]; the approaches based on BTE containing first-principles e-ph/e-d scattering rates usually applied diffusive e-b scattering[21,62], which assumes the carrier are scattered into different final states with equal possibilities and the information of initial state or boundary detail is totally lost.

A more reasonable BTE-based approach for e-b scattering takes explicit boundary scattering transition probability $P_{nn'}(\mathbf{k},\mathbf{k}')$ into account. The $P_{nn'}(\mathbf{k},\mathbf{k}')$ quantifies the transition probability for an electron from bulk initial state $n\mathbf{k}$ being scattered by boundary to bulk final state $n'\mathbf{k}'$. The classical Fuchs-Sondheimer (FS) model for electron-surface scattering widely used in 3D metal films assumes that $p$ percentage of carriers have specular scattering (i.e. $P((k_x,k_y,k_z),(k_x,k_y,-k_z)) = p$, $z$ is the norm of film surface) and the other 1-$p$ carriers have diffusive surface scattering. The empirical parameter $p$ in FS model is called surface scattering specularity and is mostly extracted from experiments. The similar empirical parameter (reflection coefficient $R$) exists in classical Mayadas-Shatzkes model for grain boundary scattering (i.e. $P((k_x,k_y,k_z),(k_x,k_y,-k_z)) = R$, $z$ is norm of grain boundary). However, the computational approach to evaluate $p$ and $R$ is still unclear. Here we stress the possibility of introducing accurate e-b scattering into the linearized BTE as boundary conditions (demonstrated in Section 2), once the e-b transition probability $P_{nn'}(\mathbf{k},\mathbf{k}')$ is known.

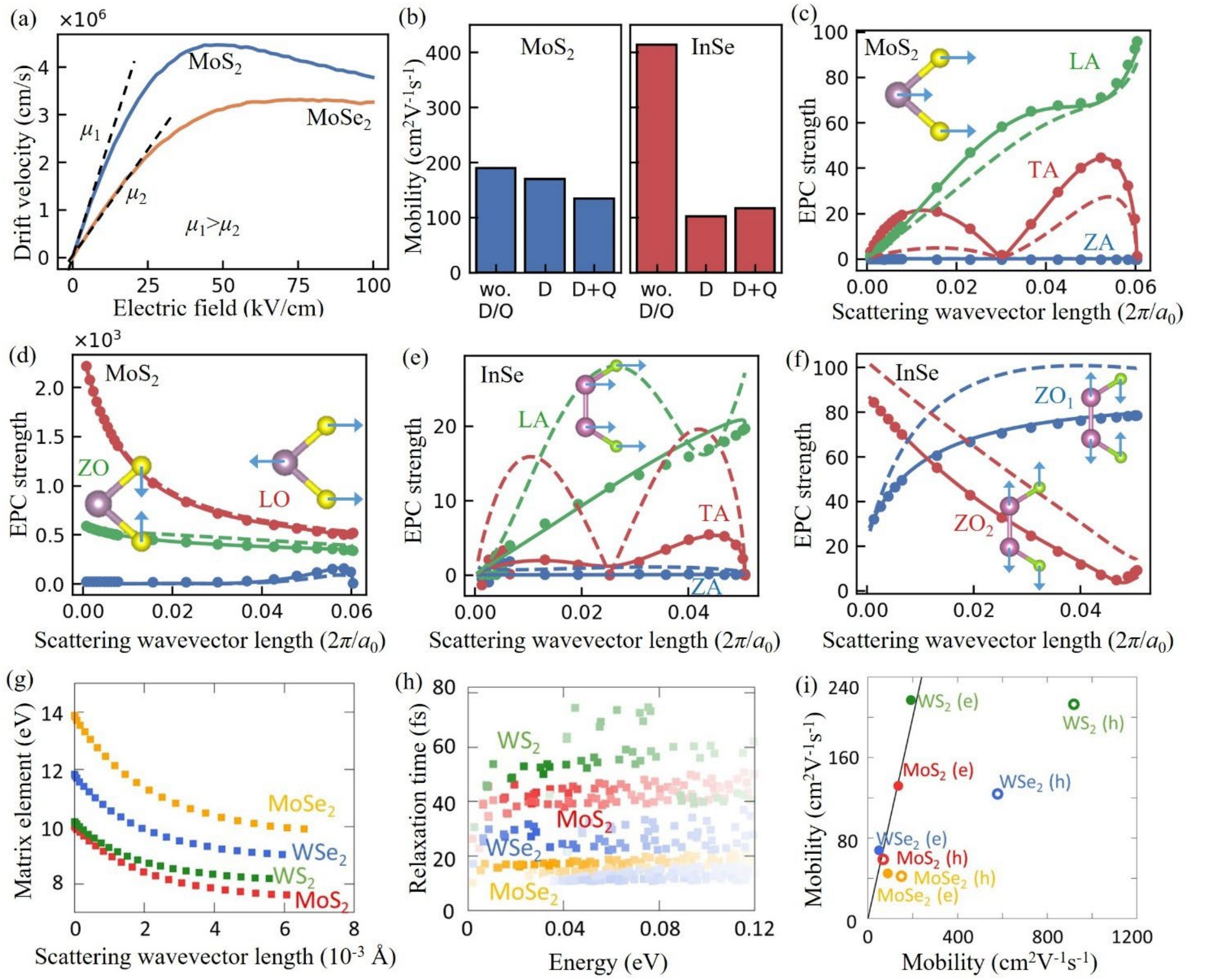

**Figure 2** (a) Phonon-limited electron drift velocity in $MoS_2$ and $MoSe_2$ under external electric field. (b) Phonon-limited electron mobility of $MoS_2$ and InSe, calculated without dipoles or quadrupoles (denoted as "wo. D/Q"), only with dipoles ("D") and with dipoles and quadrupoles ("D+Q"). (c)-(f) Comparison of the DFPT-calculated EPC strengths (dots) and the Wannier-interpolated ones (dashed lines: without quadrupoles; solid lines: with quadrupoles), for selected phonon modes of $MoS_2$ and InSe. The vibration patterns are shown in the insets. (g) Electron-defect (S/Se vacancy) interaction matrix elements for electrons in $MoS_2$, $MoSe_2$, $WS_2$ and $WSe_2$. For all materials in (c)-(g), the initial states are selected to be 30(25) meV above the CBM along the Γ-M direction, and the final states are located at the isoenergy circle of the same valley. (h) Electron relaxation time in $MoS_2$, $MoSe_2$, $WS_2$ and $WSe_2$ containing S/Se vacancy due to electron-defect scattering. (i) Defect-limited mobility (*y* axis) vs. phonon-limited mobility (*y* axis) for electron and hole in $MoS_2$, $MoSe_2$, $WS_2$ and $WSe_2$, with S/Se vacancy. The defect concentrations in (h) and (i) are $10^{12}$ $cm^{-2}$. The data in Figure 2 are from Ref. [defect xx, xx].

## 4. Electron scattering models

In this section, we briefly demonstrate the established e-ph scattering models in 2D semiconductors. Although these over-simplified models are not reliable for accurate mobility prediction, they provide fundamental understanding on e-ph scattering mechanisms and convenient estimation methods for mobility screening (see Section 6). In addition, the various **q**-dependence of EPC *g* matrix, and the corresponding energy-dependence of scattering rates are discussed and compared (see Figure 3 and Table 1), which is helpful to determine the dominant scattering from scattering rates in 2D semiconductors.

The e-ph scattering models can be classified according to perturbation potential and involved phonons. Around zone center (i.e. **q**=0), the phonon perturbation potential $V$ can be roughly decomposed to a macroscopic long-range potential $V^{\mathrm{LR}}$, which is electrostatic Coulomb potential generated by polarization due to phonon vibration and is uniform across the unit cell, and a short-range contribution $V^{\mathrm{SR}}$, which has large variation in the unit cell but is spatially periodic over unit cells. According to the type of perturbation potential ($V^{\mathrm{LR}}$ or $V^{\mathrm{SR}}$) and involved phonons (acoustic or optical phonons), e-ph scattering at long wavelength limit can be classified into four types: acoustic deformation potential scattering ($V^{\mathrm{SR}}$ induced by acoustic phonons), optical deformation potential scattering ($V^{\mathrm{SR}}$ induced by optical phonons), piezoelectric scattering ($V^{\mathrm{LR}}$ from acoustic phonons) and Fröhlich scattering ($V^{\mathrm{LR}}$ from optical phonons).

Before discussion on specific models, first we decompose the zero-order and first-order components in EPC strength $D$, which determines different **q**-dependence of $D$ in different scattering models. As shown in Appendix A, the $D$ can be written as:

$$D_{n'n\nu}(\mathbf{k},\mathbf{q}) = \langle \psi_{n'\mathbf{k}+\mathbf{q}} | \sum_{p\kappa\alpha} \tilde{e}_{\kappa\alpha,\nu}(\mathbf{q}) e^{i\mathbf{q}\cdot\mathbf{R}_p} \frac{\partial V_{\mathrm{KS}}^{(\mathrm{ph})}}{\partial \tau_{p\kappa\alpha}} | \psi_{n\mathbf{k}} \rangle, \tag{33}$$

where $\frac{\partial V_{\mathrm{KS}}^{(\mathrm{ph})}}{\partial \tau_{p\kappa\alpha}}$ is the KS potential perturbation due to atom $\kappa$ in $p$-th unit cell moving in $\alpha$ direction, $\mathbf{R}_p$ is the spatial coordinates of $p$-th unit cell, $\tilde{e}_{\kappa\alpha,\nu}$ is the atomic displacement in phonon $\nu\mathbf{q}$ defined as $\tilde{e}_{\kappa\alpha,\nu}(\mathbf{q}) = e_{\kappa\alpha,\nu}(\mathbf{q})\sqrt{M_0 / M_\kappa}$ =and $e_{\kappa\alpha,\nu}$ is the eigenvector of dynamical matrix. The $\tilde{\mathbf{e}}_\nu(\mathbf{q}) e^{i\mathbf{q}\cdot\mathbf{R}_p}$ describes how each atom moves in the crystal. At zone center, $e^{i\mathbf{q}\cdot\mathbf{R}_p} \approx 1 + i\mathbf{q}\cdot\mathbf{R}_p$ within first order of $|\mathbf{q}|$, and thus the phonon vibration at long-wavelength limit can be approximated by a periodic atomic displacement in each unit cell ($\tilde{\mathbf{e}}_\nu(\mathbf{q})$) and a lattice dilation term ($\tilde{\mathbf{e}}_\nu(\mathbf{q})\mathbf{q}\cdot\mathbf{R}_p$). By assuming initial state and final state are close enough, the EPC strength (Eq. 28) can be written as:

$$D_{n\mathbf{k},\nu} = D_{n\mathbf{k},\nu}^{(0)} + i|\mathbf{q}| D_{n\mathbf{k},\nu}^{(1)}(\hat{\mathbf{q}}), \tag{34}$$

where zero-order $D^{(0)}$ and first-order $D^{(1)}$ are:

$$D^{(0)}_{n\mathbf{k},\nu} = \langle \psi_{n\mathbf{k}} | \sum_{p\kappa\alpha} \tilde{e}_{\kappa\alpha,\nu}(\hat{\mathbf{q}}) \frac{\partial V^{\text{(ph)}}_{\text{KS}}}{\partial \tau_{p\kappa\alpha}} | \psi_{n\mathbf{k}} \rangle,$$
$$D^{(1)}_{n\mathbf{k},\nu} = \langle \psi_{n\mathbf{k}} | \sum_{p\kappa\alpha} \tilde{e}_{\kappa\alpha,\nu}(\hat{\mathbf{q}}) \hat{\mathbf{q}} \cdot \mathbf{R}_p \frac{\partial V^{\text{(ph)}}_{\text{KS}}}{\partial \tau_{p\kappa\alpha}} | \psi_{n\mathbf{k}} \rangle. \quad (35)$$

### 4.1. Acoustic deformation potential (ADP) scattering

The ADP scattering is one of the most common e-ph scatterings in carrier transport. For longitudinal acoustic (LA) phonons, the atoms in unit cell move in the same direction with same amplitudes ( $\tilde{\mathbf{e}}_\nu(\mathbf{q}) = \hat{\mathbf{q}}$ ). Therefore, the crystal undergoes a translation for zero-order $D^{(0)}$ while a unform dilation along direction of $\mathbf{q}$ ( $\sim \hat{\mathbf{q}} \cdot \mathbf{R}_p$ ) for first-order $D^{(1)}$. The $D^{(0)}$ for acoustic phonons is zero, as the translation of materials does not lead to potential perturbation for electrons in adiabatic approximation. Thus, the leading term of EPC strength for LA phonons is the first-order term which is proportional to the $|\mathbf{q}|$ (see Eq. 34). By selecting $n\mathbf{k}$ as the band edge electronic state (CBM or VBM) and focusing on isotropic semiconductors, the EPC strength for LA mode can be written as:

$$|D_{\text{LA}}(\mathbf{q})| = |\mathbf{q}| D^{(1)}_{\text{LA}}, \quad (36)$$

where $D^{(1)}_{\text{LA}}$ is the LA deformation potential constant. From Eq. 35, the $D^{(1)}_{\text{LA}}$ can be interpreted as the electronic state energy shift due to the lattice dilation, and thus can be directly calculated from regular DFT calculations on unit cell:

$$D^{(1)}_{\text{LA}} = \frac{\Delta \varepsilon_{c,v}}{\Delta l_0 / l_0}, \quad (37)$$

where $\Delta\varepsilon_{c,y}$ is the shift of CBM or VBM electronic state energy and $\Delta l_0/l_0$ is the relative deformation of lattice constant along transport direction.

Combining $D^{(1)}_{\text{LA}}$, LA sound velocity $v^{\text{LA}}$ and effective mass $m^*$, the ADP scattering rate $1/\tau_{\text{ADP}}$ is[63]:

$$\frac{1}{\tau_{\text{ADP}}(\varepsilon)} = \frac{m^* k_{\text{B}} T (D^{(1)}_{\text{LA}})^2}{\hbar^3 \rho v^2_{\text{LA}}}, \quad (38)$$

where $\rho$ is the area density and $k_{\text{B}}T$ is the temperature, which is assumed to be much larger than involved acoustic phonon energy (i.e. $k_{\text{B}}T \gg \hbar\omega$). The $\tau_{\text{ADP}}$ is independent of initial state electronic energy, diagramed in Figure 3a. As shown in Figure 3b, the 2D semiconductor $ZrI_2$ shows a dominant ADP scattering, which is indeed nearly a constant to energy. The energy-dependence of $\tau_{\text{ADP}}$ can be understood from $\mathbf{q}$-dependence of transition probability $T$ (see Eq. 10 for definition). From Eq. 10, $T$ in intrinsic semiconductor (i.e. $f_n(\mathbf{k}) \ll 1$) is proportional to product of phonon occupation $n$, square of vibration amplitude $l^2$ and square of EPC strength $D^2$. For acoustic phonon, the phonon dispersion $\omega(\mathbf{q})$ is linear with $|\mathbf{q}|$ and thus the $n$ at room temperature can be approximated by $k_{\text{B}}T / (\hbar\omega)$ for small $\omega$ and $|\mathbf{q}|$. Considering $n \sim 1/|\mathbf{q}|$, $l^2 \sim 1/|\mathbf{q}|$ and $D^2_{\text{LA}}(\mathbf{q}) \sim |\mathbf{q}|^2$, the $T$ of ADP scattering is a constant function of $|\mathbf{q}|$. From definition of MRTA scattering rate in Eq. 22, the $1/\tau_{\text{ADP}}$ is thus proportional to density of state, which is indeed a constant to energy for 2D semiconductor with a parabolic band structure.

With constant $\tau_{ADP}$, the corresponding ADP mobility $\mu_{ADP}$ can be obtained from Takagi formula[64,65]:

$$\mu_{\mathrm{ADP}} = \frac{e\hbar^3 \rho v_{\mathrm{LA}}^2}{k_{\mathrm{B}} T (m^*)^2 (D_{\mathrm{LA}})^2}. \tag{39}$$

The necessary quantities in Eq. 39 can be easily calculated and thus are available in many 2D materials database, so the $\mu_{ADP}$ can be used as a feasible indicator in high-mobility 2D semiconductor screening. The ADP scattering is universal for most materials and dominate e-ph scattering in both 3D semiconductor like Si and 2D semiconductor like $MoS_2$.

### 4.2. Optical deformation potential (ODP) scattering

In contrast to acoustic phonons, the optical phonon has nonzero zero-order EPC strength as its opposite atom movement in the unit cell leads to finite perturbation potential at **q**=0. Therefore, the optical deformation potential $D_{\mathrm{OP}}^{(0)}$ (zero-order deformation potential constant) relates to EPC strength as:

$$|D_{\mathrm{OP}}(\mathbf{q})| = D_{\mathrm{OP}}^{(0)}. \tag{40}$$

The optical deformation potential scattering rate is[66]:

$$\frac{1}{\tau_{\mathrm{ODP}}(\varepsilon)} = \frac{m^* D_{\mathrm{OP}}^2}{2\hbar^2 \rho \omega_{\mathrm{OP}}} [n_{\mathrm{OP}} + (n_{\mathrm{OP}} + 1)\Theta(\varepsilon - \hbar\omega_{\mathrm{OP}})], \tag{41}$$

where $\rho$ is the area density, $n_{OP}$ is the Bose occupation of the optical phonon with frequency $\omega_{OP}$ and $\Theta(x)$ denotes the Heavyside step function. It should be noted that the first term in square bracket of Eq. 41 represents the phonon absorption, and the latter one is for phonon-emission process, which is only possible for electron with energy larger than $\hbar\omega_{OP}$ above CBM. For each branch, the $\tau_{ODP}(\varepsilon)$ is independent of energy, which can be explained by $T$. For optical phonons, $\omega$ is a constant with respect to $|\mathbf{q}|$. Therefore, phonon occupation $n$, square of vibration amplitude $l^2$ and square of EPC strength $D^2$ are constant functions of $|\mathbf{q}|$, resulting to constant $\tau_{ODP}(\varepsilon)$ determined by density of states at energy $\varepsilon$. The most obvious feature of ODP scattering is a step-like scattering rate at $\varepsilon=\hbar\omega_{OP}$, as diagramed in Figure 3a. Figure 3b shows an example of $MoS_2$ which has significant ODP scattering at 50 meV above CBM.

### 4.3. Fröhlich scattering

The Fröhlich scattering corresponds to the zero-order EPC strength $D^{(0)}$ which originates from the long-range dipolar potentials generated by polarization from optical phonons. It is the dominant scattering in many 2D polar semiconductors[67] with an explicit expression of EPC $g$ matrix in Eq. 28 (assume the quadrupoles in **Z** is zero). The corresponding Fröhlich EPC strength $D^{\mathrm{F}}$ in isotropic materials can be written as:

$$D_j^{\mathrm{F}}(\mathbf{q}) = i\frac{e^2}{2\Omega_{\mathrm{uc}}}\sum_{\kappa}\frac{\hat{\mathbf{q}}\cdot\hat{\mathbf{Z}}_{\kappa}\cdot\tilde{\mathbf{e}}_{\kappa,j}(\mathbf{q})}{1+2\pi\alpha_{2\mathrm{D}}\,|\mathbf{q}|}, \tag{42}$$

where $\hat{\mathbf{q}}$ is the unit vector of $\mathbf{q}$, $\Omega_{uc}$ is the area of unit cell, $\kappa$ is the index of atom in the unit cell, $M_\kappa$ is the atomic mass; $j$ is the index of the optical phonon mode, $\tilde{\mathbf{e}}_{\kappa,j}$ is the atomic displacement in phonon $v\mathbf{q}$ defined as $\tilde{\mathbf{e}}_{\kappa,j}(\mathbf{q}) = \tilde{\mathbf{e}}_{\kappa,j}(\mathbf{q})\sqrt{M_0/M_\kappa}$ and $\mathbf{e}_{\kappa,j}$ is the eigenvector of dynamical matrix; $\hat{\mathbf{Z}}$ is the Born effective charge, and $\alpha_{2D}$ is the in-plane polarizability of the 2D crystal. Several approximations have been applied to derive the $D^{\mathrm{F}}$ in Eq. 42: (1) The in-plane 2D dielectric function is approximated by $1+2\pi\alpha_{2D}|\mathbf{q}|$, which is commonly used in 2D semiconductors[68,69]. (2) The wavefunction dependence has be neglected since the Fröhlich scattering is the intravalley scattering and the involved $|\mathbf{q}|$ is small. In addition, the $\mathbf{q}$ dependence of $\omega_{j\mathbf{q}}$ is neglected and the dispersion-less $\omega_j$ can be approximated by[16,70]:

$$\omega_j^2 = \omega_j^2(\mathbf{q}\to\infty) = \omega_j^2(\mathbf{q}=0) + \frac{e^2}{4\pi\alpha_{2\mathrm{D}}\Omega_{\mathrm{uc}}}\left(\sum_{\kappa}\frac{\hat{\mathbf{q}}\cdot\hat{\mathbf{Z}}_{\kappa}\cdot\mathbf{e}_{\kappa,j}(\mathbf{q}\to 0)}{\sqrt{M_\kappa}}\right)^2. \tag{43}$$

With dispersion-less $\omega_j$, the $\mathbf{q}$-dependence of transition probability $T$ for Fröhlich scattering is totally determined by $D^{\mathrm{F}}$. As shown in Eq. 42, the $D^{\mathrm{F}}$ is proportional to $1/(1+a|\mathbf{q}|)$ where $a$ is constant $2\pi\,\alpha_{2D}$. Given that $T\sim n\,l^2\,D^2$ and $n$, $l$ are determined by $\omega_j$, the $T$ is thus proportional to $1/(1+a|\mathbf{q}|)^2$, which shows strong $\mathbf{q}$-dependence in contrast to the $T$ in ADP or ODP scattering. The decreasing $T$ with respect to increasing $|\mathbf{q}|$ leads to decreasing Fröhlich scattering rate $1/\tau^{\mathrm{F}}(\varepsilon)$ with increasing energy $\varepsilon$, for both phonon-absorption and emission branches, as diagramed in Figure 3a. This is due to: for an initial state close to band extreme, the average $T$ is relatively larger owing to the smaller $|\mathbf{q}|$ between initial and final states on iso-energy circle, which leads to a larger $1/\tau^{\mathrm{F}}(\varepsilon)$; while for an initial state with higher energy above CBM (or with lower energy below VBM), the larger $|\mathbf{q}|$ leads to a relatively smaller $T$ and $1/\tau^{\mathrm{F}}(\varepsilon)$. In Figure 3b, the InSe shows typical Fröhlich scattering rate, which has two peaks at $\varepsilon=0$ and $\varepsilon=\hbar\omega_j$ both with decreasing trend with $\varepsilon$.

Combining $D^{\mathrm{F}}$ in Eq. 42 and $\omega$ Eq. 43, the corresponding EPC $g$ matrix (see Eq. 28) can be obtained by:

$$g_j^{\mathrm{F}}(\mathbf{q}) = i\frac{e^2}{2\Omega}\sum_{\kappa}\left(\frac{\hbar}{2M_\kappa\omega_j}\right)^{1/2}\frac{\hat{\mathbf{q}}\cdot\hat{\mathbf{Z}}_{\kappa}\cdot\mathbf{e}_{\kappa,j}(\mathbf{q})}{1+2\pi\alpha_{2\mathrm{D}}\,|\mathbf{q}|}. \tag{44}$$

With the $g^{\mathrm{F}}$ in hand, we can calculate the scattering rates via Eq. 22, and the Fröhlich mobility $\mu_{\mathrm{F}}$ via Eq. 24 under MRTA. The numerical calculation can be facilitated by the assumptions of a parabolic electronic band (with effective mass $m^*$) and multiple dispersion-less phonon modes (with frequency $\omega_j$). The necessary physical quantities in $\mu_{\mathrm{F}}$ can be easily calculated and are available in 2D materials database, which makes $\mu_{\mathrm{F}}$ easy to evaluate.

### 4.4. Piezoelectric scattering

In semiconductor lacking an inversion center, the acoustic phonon also generates macroscopic polarization and thus long-range perturbation potential ($V^{LR}$) similar to Fröhlich scattering, which is usually called piezoelectric (PZ) scattering. The piezoelectric EPC strength $D^{PZ}$ can be obtained from Eq. 29 by keeping quadrupoles $\hat{Q}$=and removing Born effective charge $\hat{Z}$=in $\mathbf{Z}(\mathbf{q})$ (Eq. 30). In contrast to Fröhlich scattering, the $D^{PZ}$ is proportional to $|\mathbf{q}|/(1+a|\mathbf{q}|)$ as $\hat{Q}$ is the first-order term in $\mathbf{Z}(\mathbf{q})$ around zone center. For acoustic phonon, the phonon dispersion $\omega(\mathbf{q})$ increases linearly with $|\mathbf{q}|$ and thus the $n$ is approximated by $k_B T/(\hbar\omega)$ at room temperature. Given that $n \sim 1/|\mathbf{q}|$, $l^2 \sim 1/|\mathbf{q}|$ and $D^{PZ}(\mathbf{q}) \sim |\mathbf{q}|/(1+a|\mathbf{q}|)$, the transition probability $T$ of PZ scattering is proportional to $1/(1+a|\mathbf{q}|)^2$, with the same $\mathbf{q}$-dependence as Fröhlich scattering. Consequently, the PZ scattering rate also shows a decreasing trend with increasing electronic energy, as shown in diagram in Figure 3a. The realistic examples include $MoS_2$, which shows obvious PZ scattering feature (see peak around band edge) in Figure 3b.

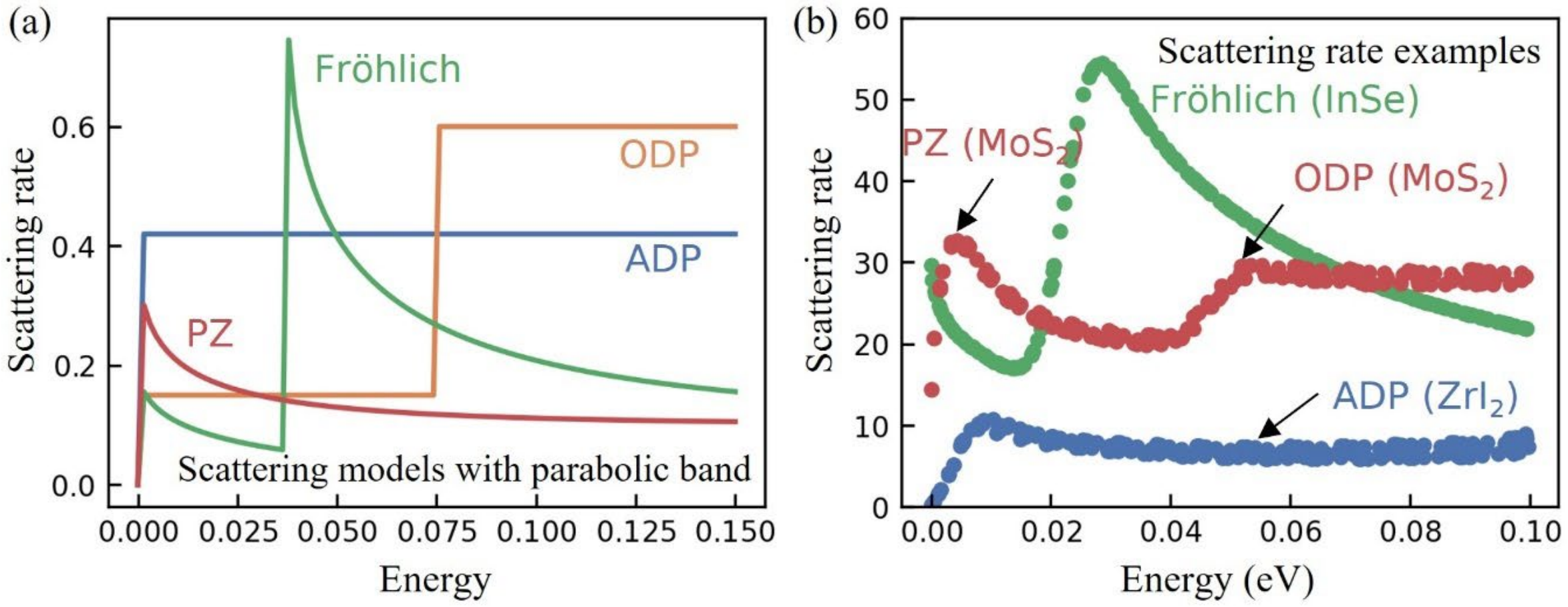

**Figure 3** (a) Model scattering rates in 2D semiconductors with parabolic band, including acoustic deformation potential scattering (ADP), optical deformation potential scattering (ODP), Fröhlich scattering (Fröhlich) and piezoelectric scattering (PZ). (b) Examples of scattering rates in realistic 2D semiconductors showing different dominant phonon scatterings.

| Scattering model | Phonon occupation ($n$) | Squared zero-point vibration amplitude ($l^2$) | Squared EPC strength ($D^2$) | Transition probability ($T$) |
|---|---|---|---|---|
| ADP | $\frac{1}{\lvert\mathbf{q}\rvert}$ | $\frac{1}{\lvert\mathbf{q}\rvert}$ | $\lvert\mathbf{q}\rvert^2$ | 1 |
| ODP | 1 | 1 | 1 | 1 |
| PZ | $\frac{1}{\|\mathbf{q}\|}$ | $\frac{1}{\|\mathbf{q}\|}$ | $\frac{\|\mathbf{q}\|^2}{(1+a\|\mathbf{q}\|)^2}$ | $\frac{1}{(1+a\|\mathbf{q}\|)^2}$ |
| Fröhlich | 1 | 1 | $\frac{1}{(1+a\|\mathbf{q}\|)^2}$ | $\frac{1}{(1+a\|\mathbf{q}\|)^2}$ |

**Table 1** Summary of **q**-dependence of phonon occupation $n$, squared zero-point vibration amplitude $l^2$, squared EPC strength $D^2$ and transition probability $T$ for different e-ph scattering models in 2D semiconductors.

## 5. Dimensional effect in mobility: a prospective from "density of scattering"

One of the grand challenges for electronic materials research is to find an alternative to silicon with a suitable band gap, high carrier mobility at room temperature, and ambient stability, when thinning down to atomic thickness. Since the discovery of graphene in 2004, various of 2D materials have been synthesized and proposed as candidate materials in 2D transistors but none of them is satisfactory. For example, although graphene has very high carrier mobility, it does not have band gap. Some 2D semiconductors, including $MoS_2$, $MoSe_2$, $WS_2$, $WSe_2$, have moderate band gap but generally suffer from low carrier mobility at room temperature. Although it is reported that high carrier mobility (~ 1000 $cm^2V^{-1}s^{-1}$) can be achieved in 2D semiconductors like phosphorus[71], InSe[72], the thickness of measured channel materials is around 10 nm, which violates the purpose of transistor scaling-down.

Indeed, despite extensive research on monolayer semiconductors, their mobilities are currently far below Si (~ 1400 $cm^2V^{-1}s^{-1}$) due to strong e-ph and/or e-d scattering. In Figure 4 and Table 2 in Appendix D, we review the experimentally reported mobilities of existing monolayer semiconductors (bright colors). $WSe_2$ holes have the highest mobility (500-655 $cm^2V^{-1}s^{-1}$) as reported in a very recent study[14], while all the others have mobility < 200 $cm^2V^{-1}s^{-1}$. The data of multilayer semiconductors with thickness from several to tens of nanometers is also included in Figure 4 and Table 2 for comparison. It can be found that increasing the thickness indeed increases the mobility, such as InSe and black phosphorus multilayers.

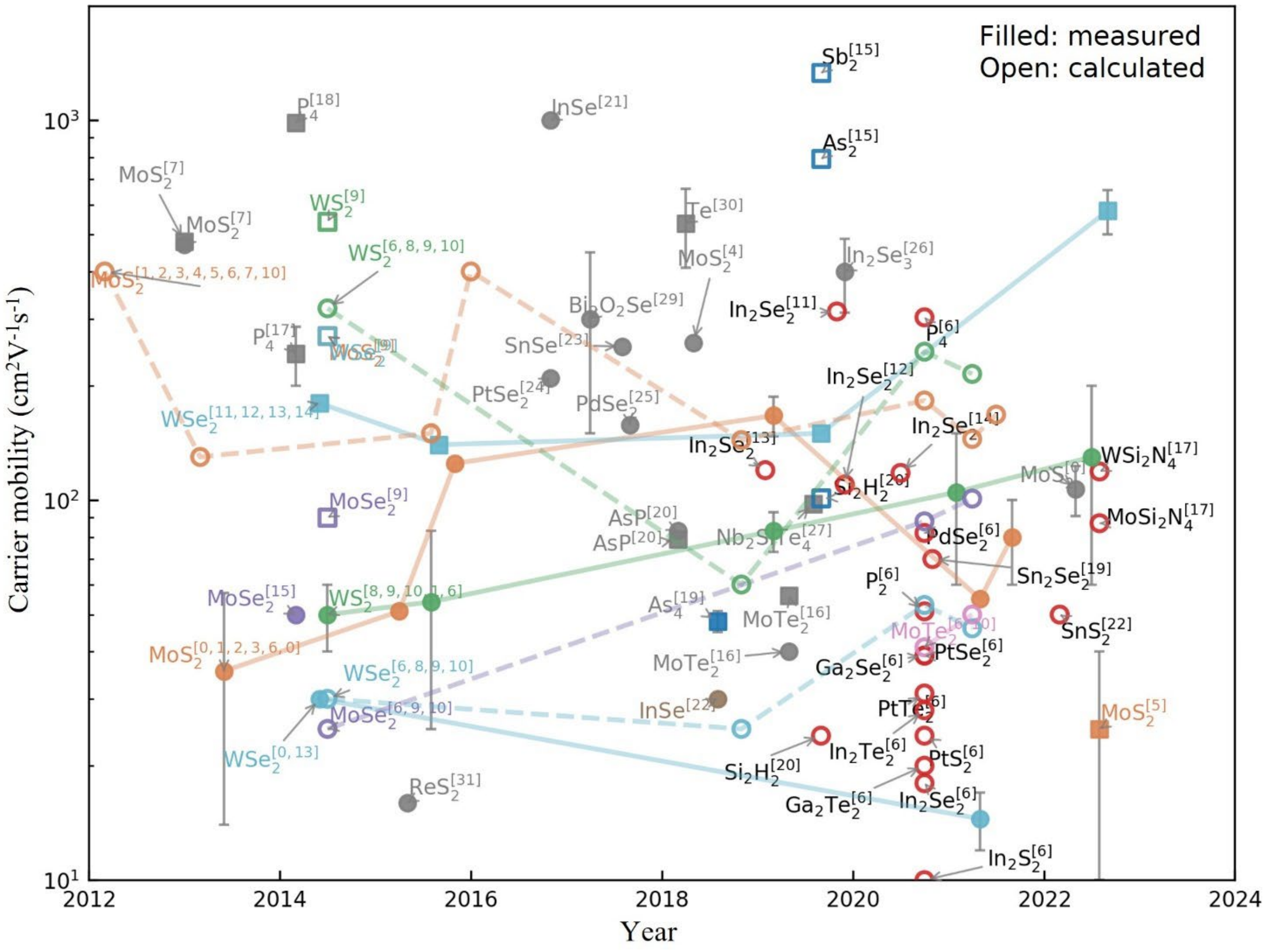

**Figure 4** Review of experimental (filled symbols) and computational (open symbols) works on carrier mobility of 2D semiconductors. The monolayers are shown in bright colors while the multilayers (ranging from several to tens of nanometers) are shown in grey. Electrons are marked by circles while holes are by squares. The $MoS_2$ electrons, $WS_2$ electrons and $WSe_2$ holes repeatedly reported in different papers over the years are connected with lines. The experimental data are from Refs. [14,52,71-104] and computational date are from Refs. [18,25,35,105-122] (also see Appendix D for values).

The universally lower carrier mobility in 2D semiconductors is believed to originate from the dimensional effect[113,115]: for a parabolic electronic band structure, reducing the dimension will increase the density of states near the band edge; similarly, for a linear phonon band, reducing the dimension will also increase the density of low-energy phonons; therefore, 2D semiconductors tend to have a higher density of scatterings and thus a lower mobility. In the following, we use $MoS_2$ and bulk silicon (Si) as two representative semiconductors for 2D and 3D to further demonstrate the dimensional effect in carrier mobility.

First, the carrier mobility of $MoS_2$ and Si are decomposed into their corresponding "Drude effective mass" ($\bar{m}^*$) and "Drude scattering rate" ($1/\bar{\tau}$), which are defined by re-writing the mobility $\mu$ (defined in Eq. 24) in a form similar to the Drude model:

$$\mu = \frac{|q|\bar{\tau}}{\bar{m}^*}, \tag{45}$$

where

$$\frac{1}{\bar{m}^*} = \frac{1}{n_e \Omega N_d} \sum_n \int \frac{d\mathbf{k}}{\Omega_{\mathrm{BZ}}} \frac{\partial f_n^0(\mathbf{k})}{\partial \varepsilon_{n\mathbf{k}}} v_{n\mathbf{k}}^2, \qquad \bar{\tau} = \frac{\mu \bar{m}^*}{|q|}. \tag{46}$$

$N_d$ is the dimensions of $MoS_2$ and Si. The $\bar{m}^*$ is fully determined by the electronic structure and its occupation, while the information about e-ph/e-d scattering are wrapped in $1/\bar{\tau}$. The scalar $\mu$ is reduced from mobility tensor in Eq. 24 by taking the average of diagonal term $\mu = \sum_\alpha \mu_{\alpha\alpha}$, as $MoS_2$ ($\alpha$ runs in $x$ and $y$) and Si ($\alpha$ runs in $x$, $y$ and $z$) both are isotropic materials. In intrinsic semiconductors, it can be found that the $\bar{m}^*$ would be very close to band edge effective mass by noticing that:

$$\begin{aligned}\frac{1}{\bar{m}^*} &= \frac{1}{n_e \Omega N_d} \sum_{n\alpha} \int \frac{d\mathbf{k}}{\Omega_{\mathrm{BZ}}} v_{n\mathbf{k},\alpha} \frac{\partial f_n^0(\mathbf{k})}{\partial \hbar k_\alpha} \\ &= \frac{1}{n_e \Omega N_d} \sum_{n\alpha} \int \frac{d\mathbf{k}}{\Omega_{\mathrm{BZ}}} m_{n\mathbf{k},\alpha\alpha}^{-1} f_n^0(\mathbf{k}),\end{aligned} \tag{47}$$

where $m_{n\mathbf{k},\alpha\alpha}^{-1} = \frac{\partial v_{n\mathbf{k},\alpha}}{\partial \hbar k_\alpha}$ is the state-resolved inverse effective mass tensor. The $\mu$, $\bar{m}^*$ and $1/\bar{\tau}$ of $MoS_2$ and Si are compared in Figure 5a. Here we report the Si electron mobility as 1401 $cm^2V^{-1}s^{-1}$ calculated on an equal footing with the $MoS_2$ (i.e. the electron-phonon coupling (EPC) matrix elements are obtained from DFPT + interpolation with quadrupole scattering), which is consistent with experiments[123] and other first-principles calculations[22,124]. The electron mobility of $MoS_2$ is 134 $cm^2V^{-1}s^{-1}$, which is only tenth of that in Si. The large mobility difference cannot be accounted by the 50% difference in $\bar{m}^*$ (0.43 $m_e$ in $MoS_2$ and 0.29 $m_e$ in Si) but is largely determined by their difference in $1/\bar{\tau}$ (30 $ps^{-1}$ in $MoS_2$ vs. 4.4 $ps^{-1}$ in Si).

In order to fully understand the $1/\bar{\tau}$ difference, we further calculate EPC $g$ matrix for representative EPCs and the mode-resolved "density of scatterings" $D^{\mathrm{S}}$, which has a form very similar to the scattering rate defined in Eq. 22:

$$\begin{aligned}D_{\nu,n\mathbf{k}}^{\mathrm{S}} = \frac{2\pi}{\hbar N_{\mathbf{k}'}} \sum_{n'\mathbf{k}'} &[(f_{n'}^0(\mathbf{k}') + n_{\nu\mathbf{q}})\delta(\varepsilon_{n\mathbf{k}} - \varepsilon_{n'\mathbf{k}'} + \hbar\omega_{\nu\mathbf{q}}) \\ &+ (1 + n_{\nu\mathbf{q}} - f_{n'}^0(\mathbf{k}'))\delta(\varepsilon_{n\mathbf{k}} - \varepsilon_{n'\mathbf{k}'} - \hbar\omega_{\nu\mathbf{q}})](1 - \cos\theta_{n\mathbf{k},n'\mathbf{k}'}),\end{aligned} \tag{48}$$

except that the EPC $g$ matrix is not contained and the phonon modes are not summed. The $D^{\mathrm{S}}$ can be interpreted as the density of scattering events for an initial state under energy and momentum conservation, which is fully determined by the "match" between electron and phonon band structures as well as their occupations. A higher sound velocity would reduce the phonon populations, and thus lower the $D^S$, thereby resulting in a lower scattering rate and hence a higher carrier mobility. Similarly, a smaller effective mass would reduce the density of electronic states, which leads to a lower $D^S$.

In Figures 5c-e, the LA energy-resolved scattering rate, $D^S$ and representative $|g|^2$ for $MoS_2$ and Si are compared. For 2D $MoS_2$, we consider the $|g|^2$ between the initial state located at 30 meV above the CBM along Γ-M direction, and the final states at the iso-energy circle of the same valley. For 3D Si, as there are 6 equivalent CBMs, so we select one initial state 30 meV above CBM and 100 final states evenly distributed on the iso-energy surface in the same valley assuming an anisotropic parabolic band (see Figure 5b for their distribution in the BZ). The $|g|^2$ between these initial and final states for $MoS_2$ and Si are shown in Figure 5e. As we can see, Si has larger LA $|g|^2$ than $MoS_2$, due to the larger ADP constant in Si (8.21 eV for Si[124] and 6.86 eV for $MoS_2$[125]) and larger available scattering wavevector length $|\mathbf{q}|$. However, in spite of the larger $|g|^2$, the Si has much lower scattering rate than $MoS_2$, due to its negligible $D^S$ (see Figure 5d). The remarkably lower $D^S$ in bulk Si cannot be explained by the difference of LA sound velocity (8.5 km/s in Si and 6.6 km/s in $MoS_2$) or effective mass (0.29 $m_e$ in Si an 0.43 $m_e$ in $MoS_2$), and thus is attributed to the dimensional difference of the BZ integrated in Eq. 48.

The larger acoustic phonon $D^S$, leading to larger scattering rate and lower mobility, is believed to be universal for all 2D semiconductors[113]. Indeed, the LA $D^S$ of more 2D semiconductors are shown in Figure 8e and all of them are significantly above that of Si. Since the acoustic phonon universally exists in materials, the phonon-limited mobility degradation in 2D semiconductors due to dimensional effect in $D^S$ seems inevitable. For e-d scattering, the mobility degradation is also expected to exist, where the $D^S$ is defined as:

$$D^S_{n\mathbf{k}} = \frac{2\pi}{\hbar N_{\mathbf{k}'}} \sum_{n'\mathbf{k}'} \delta(\varepsilon_{n\mathbf{k}} - \varepsilon_{n'\mathbf{k}'})(1 - \cos\theta_{n\mathbf{k},n'\mathbf{k}'}). \quad (49)$$

The definition of e-d scattering $D^S$ in Eq. 49 resembles that of density of states, which is independent of energy $\varepsilon$ in 2D semiconductor while proportional to $\sqrt{\varepsilon}$ in 3D semiconductor. Therefore, at band edge, the $D^S$ of e-d scattering in 2D semiconductor is supposed to be larger, leading to a lower defect-limited mobility if other factors are fixed.

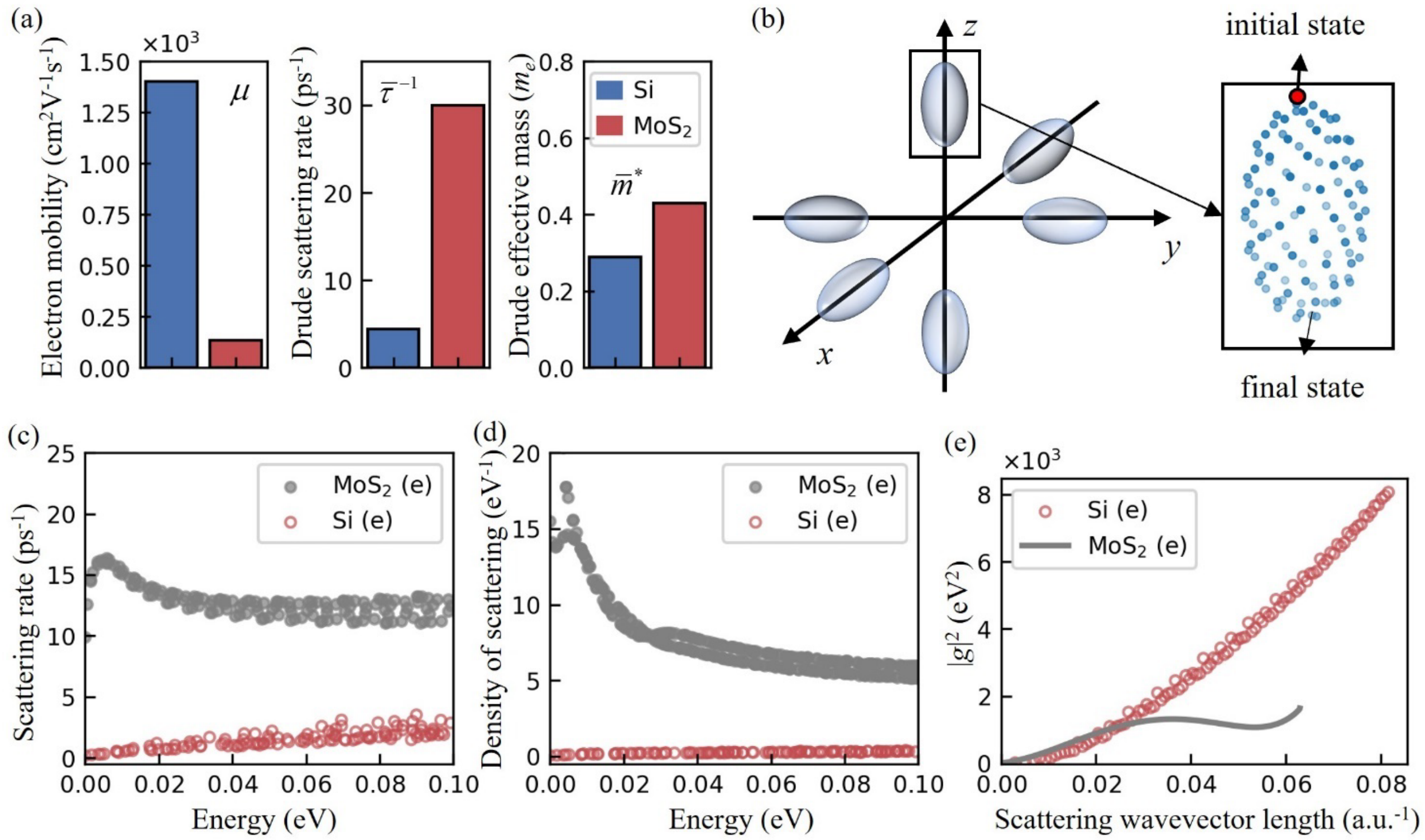

**Figure 5** (a) Comparison of bulk silicon and 2D $MoS_2$ monolayer, for electron mobility, Drude scattering rate and Drude effective mass. (b) Diagram of six conduction band valleys in bulk silicon. (c)-(e) LA-phonon-mode-resolved scattering rates (c), density of scatterings (d) and representative (squared) EPC $g$ matrix (e) for silicon and $MoS_2$ electrons. For both materials, the initial states are selected to be 30 meV above the CBM, and the final states are located at the iso-energy circle (iso-surface for silicon; diagramed in b) of the same valley.

## 6. High-mobility 2D semiconductor screening

As reviewed in Section 3, the first-principles methods developed for e-ph scattering facilitate the accurate calculation of phonon-limited mobility for 2D semiconductors. Although the dimensional effect of $D^S$ leads to mobility degradation in 2D semiconductors, there are still several high first-principles mobilities (> 1000 $cm^2V^{-1}s^{-1}$) reported in recent years. In Figure 4, we reviewed the computationally reported phonon-limited mobilities based on first-principles EPC $g$ matrix for 2D semiconductors. The room temperature hole mobility of $Sb_2$ and electron mobility of $Ge_2H_2$ are predicted to be 1330 and 2380 $cm^2V^{-1}s^{-1}$ respectively[109,117], revealing the possibility of high mobility in 2D semiconductors. However, the previous studies focused on case materials, which only explored small percentage of all 2D semiconductors. Thorough and systematical study covering more materials is desirable for fully discovering and better understandings of high mobilities in 2D semiconductors. In the following, we focus on two recent high-throughput computation on 2D semiconductor phonon-limited mobility, which screened 14 high mobilities (>1400 $cm^2V^{-1}s^{-1}$) in total (also see Figure 7 for full results)[15,16].

The basic idea of high-throughput carrier mobility calculation is to benefit from well-established materials database, select a series of 2D semiconductor with desirable properties for applications (e.g., moderate bandgap, stability, effective mass) and accurately calculate the carrier mobility using the BTE in conjunction with DFPT. The calculation results provide not only the lists of high-mobility 2D semiconductors candidates, but also the large number of examples for in-depth analysis and understanding. By connecting the basic physical properties with the calculated carrier mobility, the dominant factors in e-ph scattering and "genome" of high mobilities can be determined, which facilitates further high-mobility 2D semiconductor discovering.

The first step in high-throughput mobility calculation is to screen the potential candidates for accurate EPC calculations from 2D materials database. The well-established 2D materials databases include the Computational 2D materials Database (C2DB)[126,127] and the Materials Cloud 2D crystals database (MC2D)[128], which contains 4000 and 3000 2D materials respectively. The numerous amounts of 2D semiconductor in database and the large computational cost for EPC calculation make the accurate mobility calculations for all semiconductors prohibited. Different screening approaches are used in Refs 15 and 16. In Ref. 15, a heuristic approach is used to narrow down the computation candidates. Inspired by InSe and $P_4$, single-valley 2D semiconductors (band gap < 2.5 eV) with large Fermi velocity (maximum Fermi velocity $v_{\max} > 6$ ARU), and those with decent Fermi velocity ($v_{\max} >$ 2 ARU) but high velocity ratio ($v_{\max} > 1.7\ v_{\min}$) are screened out from the MC2D for accurate calculations, generating 11 2D semiconductors with potential high mobilities. In Ref. 16, several descriptors with physical knowledge of scattering are used in screening, including "combined effective mass" $M$, acoustic deformation potential mobility $\mu_{\mathrm{ADP}}$ and Fröhlich mobility $\mu_{\mathrm{F}}$. The $\mu_{\mathrm{ADP}}$ and $\mu_{\mathrm{F}}$ are introduced in Sections 4.1 and 4.3. The $M$ is defined as:

$$M = \sqrt{m_{\mathrm{t}}^* m_d^*}, \tag{50}$$

where $m_{\mathrm{t}}^*$ is the effective mass along carrier transport direction, and $m_{\mathrm{d}}^*$ is the density of state effective mass that can be approximated by $N\sqrt{m_x^* m_y^*}$ (here $N$ is the degeneracy of conduction/valence band extremes, and $x$ and $y$ are the two directions defined in the database). The $M$ include the information of both $m_{\mathrm{d}}^*$ and $m_{\mathrm{t}}^*$ because: (1) A lower $m_{\mathrm{d}}^*$ indicates a lower density of electronic states and thus less states available for carriers to be scattered to, which can increase the mobility as exemplified[117,129] by $Sb_2$ and $WS_2$. (2) When the scattering is fixed, decreasing $m_{\mathrm{t}}^*$ can further improve the mobility according to the Drude model.

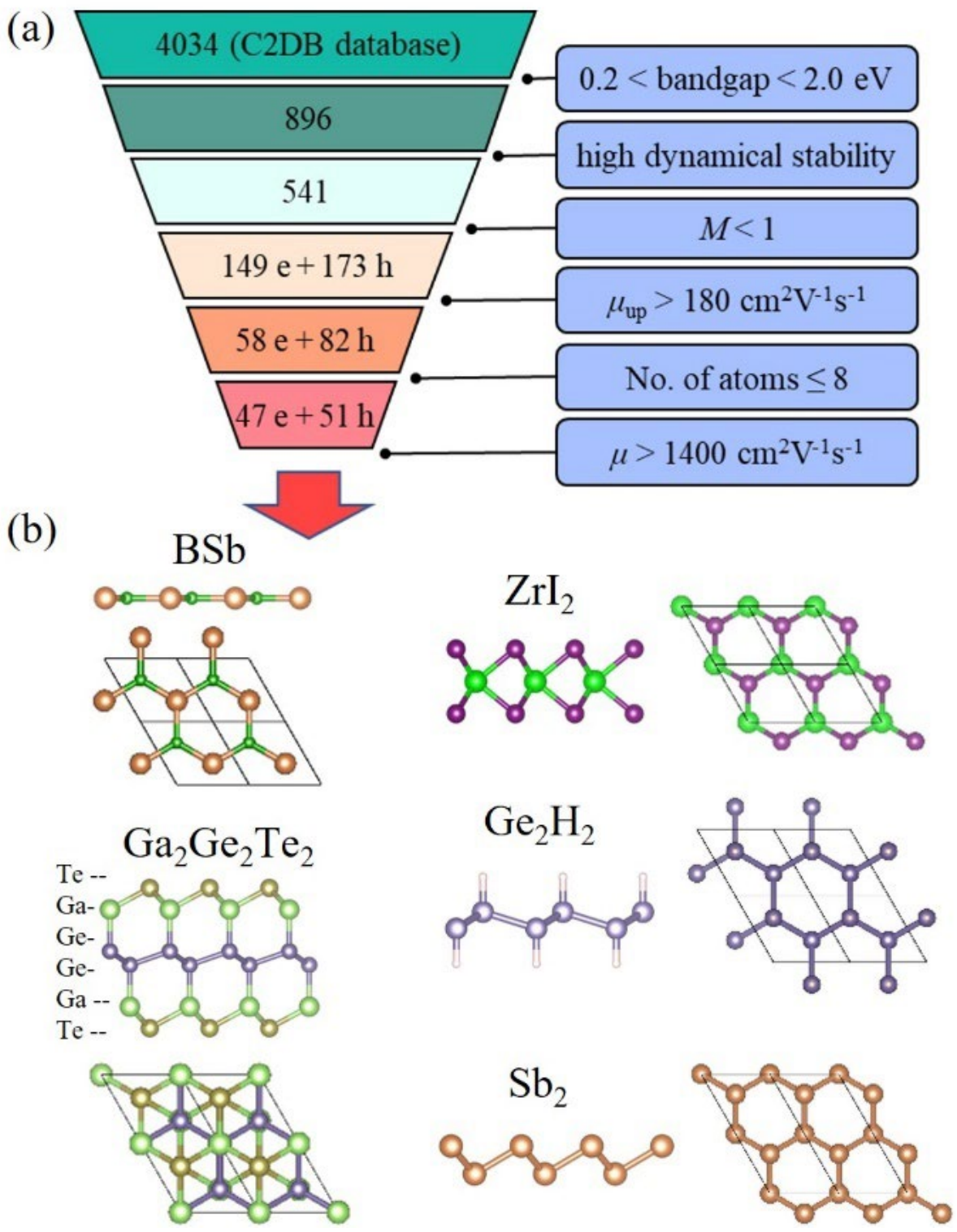

**Figure 6** (a) Screening procedures to discover 2D semiconductors with potential high carrier mobility in Ref. 16. (b) Crystal structures for representative 2D semiconductors with mobilities over 1400 $cm^2$ $V^{-1}$ $s^{-1}$ in both Refs 15 and 16.

The screening steps in Ref. 16 are illustrated in Figure 6a. First, the 2d materials with PBE band gap within 0.2-2 eV are extracted from the C2DB and then those with "high dynamical stability" are reserved for further screening. The dynamical stability in C2DB is judged from two conditions: all phonon modes having positive frequencies and elastic stiffness tensor having positive eigenvalues, which together indicates the unstressed 2D crystal is stable with no external loads in the harmonic approximation[127,130]. Then the remaining 541 out of 4000 2D materials are screened by $M$<1, referring to $M$=0.85 in $MoS_2$. Since the $M$ has different value for conduction/valence band, the $M$ criterion gives 149/179 materials for electron/hole transport. Then two more quantitative descriptors $\mu_{ADP}$ and $\mu_F$ calculated from scattering models (see Sections 4.1 and 4.3) are used. The upper limit mobility (denoted as $\mu_{up}$) is obtained following the Matthiessen's rule: $\mu_{up}^{-1} = \mu_F^{-1} + \mu_{ADP}^{-1}$, which gives an estimation of upper limit of real mobility. By taking $\mu_{up}$ > 180 $cm^2V^{-1}s^{-1}$ as a criterion and excluding those with more than 8 atoms in the unit cell, eventually 47 electron and 51 hole mobilities are calculated with first-principles calculations in Ref. 16.

The overall results with Refs. 15 and 16 are concluded in Figure 7. In particular, 14 materials with mobility larger than bulk Si electron (1400 $cm^2V^{-1}s^{-1}$) are found, all having hexagonal lattice and isotropic mobilities, with 5 kinds of crystal structures as shown in Figure 6b: (1) III-V materials: BSb (e: 5167; h: 6935; 'e' for electron and 'h' for hole; in the unit of $cm^2V^{-1}s^{-1}$), AlBi (e: 2835; h: 3446), GaSb (e: 1809), BAs (e: 1524; h: 2439), InN (e: 2106) and BP (h: 1921). They all have an atomically flat structure like graphene. (2) $ZrI_2$ (h: 5138), $HfI_2$ (h: 4782) and $WSe_2$ (h: 1962). Their structures are similar to that of 2H phase $MoS_2$ with the metal layer in the middle. (3) H functionalized IV materials: $Sn_2H_2$ (e: 3227; h: 2063) and $Ge_2H_2$ (e: 2791). (4) Group V materials: $Sb_2$ (h: 2044). (5) $Ga_2Ge_2Te_2$ (e: 1996) and $Al_2Ge_2Te_2$ (e: 2023). They have unique sextuple layered structure with atomic layers in order of Te-Ga(Al)-Ge-Ge-Ga(Al)-Te. Particularly, BSb, AlBi, $Sn_2H_2$ and BAs are predicted to have both high electron mobility and hole mobility, as characterized by the ambipolar mobility $\mu_a$ (defined by $\mu_a = 2\mu_e\mu_h/(\mu_e+\mu_h)$ where $\mu_e$ and $\mu_h$ are electron and hole mobility respectively): 5922 for BSb, 3111 for AlBi, 2517 for $Sn_2H_2$ and 1876 for BAs. These excellent properties make them especially promising for electronic devices.

The first-principle mobilities $\mu$ in Figure 7 not only provides a list of high-mobility 2D semiconductors, but also data to examine the quality of screening steps in Figure 6a, especially for $\mu_{\mathrm{up}}$ step. In Figure 8a, the $\mu$ and $\mu_{\mathrm{up}}$ of those calculated 2D semiconductor in Ref. 16 are compared, with a solid line indicating $\mu=\mu_{\mathrm{up}}$ and a dashed line for $\mu=1.38\mu_{\mathrm{up}}$. Indeed, the most of materials fall below the solid line, meaning $\mu<\mu_{\mathrm{up}}$. By enlarging the tolerance, 98% materials in Figure 8a have $\mu<1.38\mu_{\mathrm{up}}$. Combining the $\mu_{\mathrm{up}}>180\ \mathrm{cm^2V^{-1}s^{-1}}$ criterion used in screening step, the comparison in Figure 8a indicates that the rest of 2D materials filtered out are most likely to have a real $\mu < 180\times1.38=250\ \mathrm{cm^2V^{-1}s^{-1}}$.

It can be noticed that there is large discrepancy between the reported mobility from Ref. 15 and Ref. 16 for the same material. For example, $WSe_2$ hole mobility in Ref. 15 is 1962 $\mathrm{cm^2V^{-1}s^{-1}}$ while is 578 $\mathrm{cm^2V^{-1}s^{-1}}$ in Ref. 16. This can be mainly attributed to different carrier concentrations $n_c$ considered in Refs. 15 and 16. The Ref. 16 calculated 2D semiconductors at a low $n_c$ limit. In contrast, for Ref. 15, a high carrier concentration ($n_c = 10^{13}\ \mathrm{cm^{-2}}$) is considered within gated DFPT setup[36], which changed the Fermi level position and more importantly, affected EPC $g$ matrix. The most significant consequence of free carrier doping is the screening of piezoelectric and Fröhlich scattering, leading to the larger $WSe_2$ hole mobility calculated at high $n_c$. The free carrier doping effect on EPC $g$ matrix is actually not limited to the long-range scattering (e.g. piezoelectric and Fröhlich scattering) and its further influence on EPC will be discussed in detail in Section 7. In addition to $n_c$, the other differences in Refs. 15 and 16 include: (1) Materials exfoliation: The 2D materials from Ref. 15 are selected from those with layered 3D counterparts and low binding energy which can be extracted from MC2D. While Ref. 16 is based on C2BD which includes all potential 2D materials based on crystal structures of existing 2D prototype and thus not all materials in Ref. 16 are exfoliable monolayers, demanding other synthesis methods for high-mobility monolayers. (2). Spin-orbit coupling: The spin-orbit coupling is included in first-principles calculations of Ref. 16 but not considered in Ref. 15 except $WSe_2$. (3). Solution of BTE: Iterative solution of BTE is used in Ref. 15 but MRTA is applied in Ref. 16, which might lead to mobility deviations by tens of percentage[24,25].

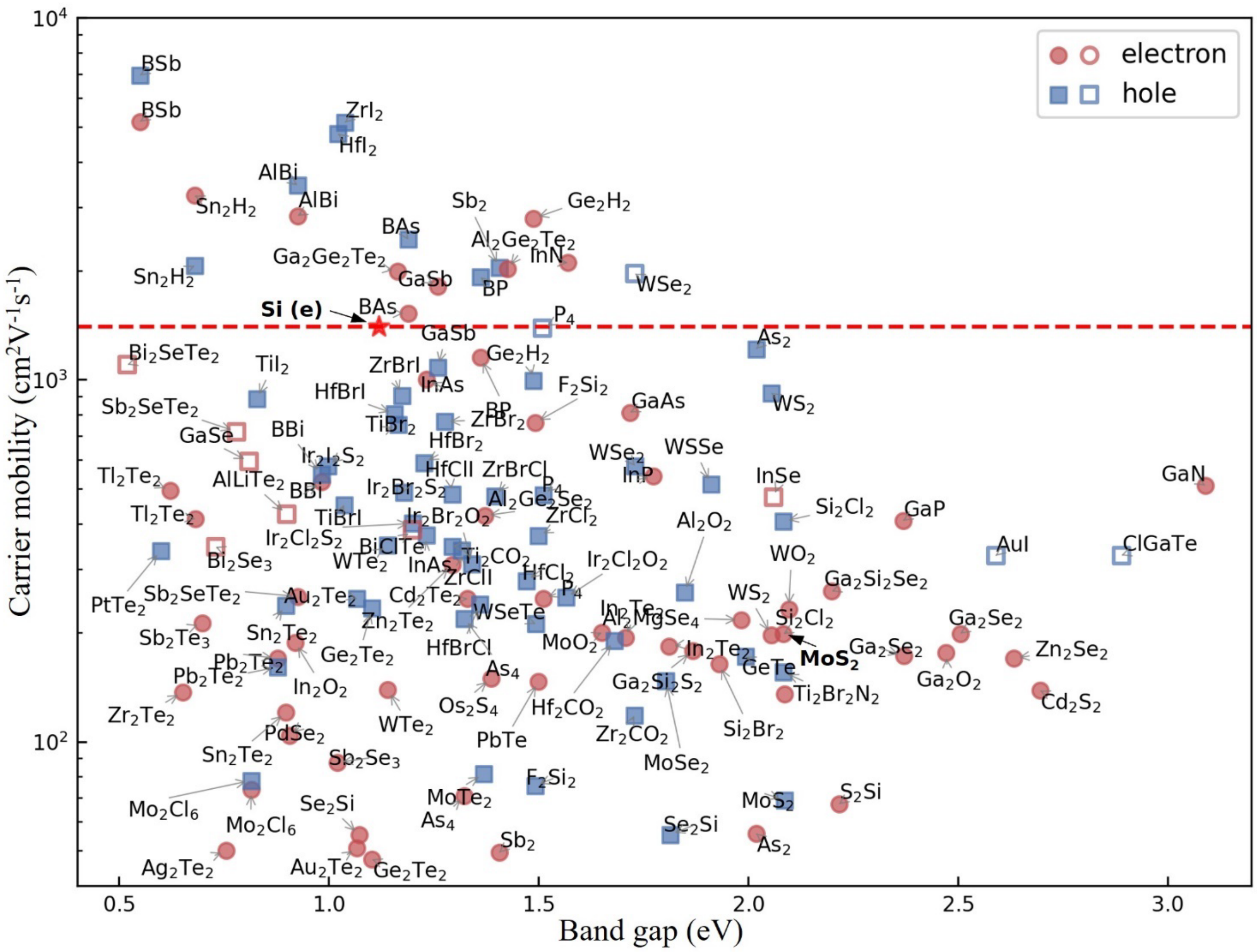

**Figure 7** Phonon-limited mobility vs. band gap (HSE) for 2D semiconductors from high-throughput calculations[15,16]. For comparison, the electron mobilities for $MoS_2$ and bulk Si are marked. The closed symbols indicate intrinsic 2D semiconductor and opened ones consider 2D semiconductor electrostatically doped at carrier concentration $10^{13}$ cm$^{-2}$.

The next step is to understand what contributes to the high mobility found in Refs. 15 and 16. In Ref. 15, two contributions to carrier mobility are focused: relaxation time $\tau$ and group velocity projection on electric field direction $v \cos\theta_u$ ($\theta_u$ is the angle between velocity and electric field). The relaxation time $\tau$ is mainly limited by the EPC strengths, which is attributed to the specific materials properties. The optimal $v \cos\theta_u$ can benefit from two types of band structures. The first one is an isotropic steep and deep valley, which enables extremely large $v$ at Fermi surface with high doping ($n_c$=$10^{13}$ cm$^{-2}$). However, in isotropic valleys, there are partial carriers moving in the perpendicular direction to electric field and thus leading to little contribution to transport (i.e. $\cos\theta_u$=0). The second type of optimal band structure has a highly anisotropic band edge ($v_{max} > 1.7\ v_{min}$), leading to more electronic states moving in field direction ($\cos\theta_u$ ~1; suppose $v_{max}$ is in direction of electric field) and thus larger $v \cos\theta_u$. The $Bi_2SeTe_2$ and $P_4$ are representatives of two optimal band structures, having 719 cm$^2$V$^{-1}$s$^{-1}$ (electron) and 1386 cm$^2$V$^{-1}$s$^{-1}$ (hole) mobility respectively.

In Ref. 16, the mobility is decomposed into 3 factors for understandings: Drude effective mass $\bar{m}^*$, density of scattering $D^S$ and EPC $g$ matrix (see Section 5 for respective definitions). Further comparisons of $\bar{m}^*$, $D^S$ and $g$ reveal that the higher mobilities found in Ref. 16 than that of $MoS_2$ and bulk Si can be attributed to small $\bar{m}^*$ and/or small $g$. In Figure 8b, the Drude effective mass $\bar{m}^*$ and Drude scattering rate $1/\bar{\tau}$ for high-mobility 2D semiconductors in Ref. 16 as well as those for $MoS_2$ and Si are shown. The $\bar{m}^*$ is fully determined by the electronic structure and its occupation, while the information about $D^S$ and EPCs are wrapped in $1/\bar{\tau}$. Most of high-mobility 2D semiconductors benefit from their low $\bar{m}^*$, compared to Si and $MoS_2$. For example, the $Ga_2Ge_2Te_2$ has a relatively large $1/\bar{\tau}$ (15 vs 30 $ps^{-1}$ in $MoS_2$ and 4.4 $ps^{-1}$ in Si) but a small $\bar{m}^*$ (0.058 vs 0.43 $m_e$ in $MoS_2$ and 0.29 $m_e$ in Si), leading to a high mobility (1996 vs 136 $cm^2V^{-1}s^{-1}$ in $MoS_2$ and 1400 $cm^2V^{-1}s^{-1}$ in Si). However, exceptions include $HfI_2$ and $ZrI_2$, which have relatively large $\bar{m}^*$ ($HfI_2$: 0.32 $m_e$; $ZrI_2$: 0.40 $m_e$) but extremely small $1/\bar{\tau}$ ($HfI_2$: 1,1 $ps^{-1}$; $ZrI_2$: 0.85 $ps^{-1}$). BSb has both a small $\bar{m}^*$ (0.09 $m_e$ for hole and electron) and a low $1/\bar{\tau}$ (2.8 $ps^{-1}$ for hole and 3.6 $ps^{-1}$ for electron), which together make it the highest mobility material.

The effects of $D^S$ and EPC $g$ matrix on carrier mobility is further displayed in Figures 8d-i, focusing on 4 representative materials: 3 high-mobility 2D semiconductor BSb, $ZrI_2$ and $Ga_2Ge_2Te_2$, and 1 common 2D semiconductor $MoS_2$ with low mobility. The mode-resolved scattering rates $\tau^{-1}$ and $D^S$ for LO and LA modes are compared across materials in Figure 8e and h and the representative $|g|^2$ between the initial state (located at 30 meV above/below the CBM/VBM along Γ-M direction) and the final states (at the iso-energy circle of the same valley) are shown in Figure 8f and i. Although $Ga_2Ge_2Te_2$ has a lower $D^S$ than $MoS_2$ for the LO mode due to its smaller effective mass (0.05 vs. 0.5 $m_e$), it has larger $|g|^2$, leading to the stronger LO scattering. The larger LO $|g|^2$ in $Ga_2Ge_2Te_2$ can be explained by its larger ratio of the Born charge to the in-plane polarizability: $R_{B/P}$=0.23 ($R_{B/P}$=0.07 in $MoS_2$) and consequently larger Fröhlich scattering (see Eq. 42). For LA mode, the $D^S$ of $Ga_2Ge_2Te_2$ is comparable to that of $MoS_2$, while the $|g|^2$ is smaller, resulting in the weaker LA scattering. In contrast to $Ga_2Ge_2Te_2$, $ZrI_2$ has a larger $D^S$ than $MoS_2$ for LO mode resulting from its smaller LO frequency (15 vs. 47 meV in $MoS_2$), however, its $|g|^2$ is smaller owing to the vanishing $R_{B/P}$ ($1.7\times10^{-4}$, which is mainly caused by the smaller Born charge: 0.0012 vs. 0.5 in $MoS_2$), causing its nearly vanishing Fröhlich scattering. The $|g|^2$ of LA mode in $ZrI_2$ is also smaller than that in $MoS_2$, giving rise to its smaller LA scattering. For BSb, although it has the largest LO $|g|^2$ because of its largest $R_{B/P}$ (0.51) among the 4 materials compared here, its LO $D^S$ is extremely small due to the highest LO phonon frequency (88 meV), and thus it has a weak LO scattering. BSb also has a lower $D^S$ for LA mode than $MoS_2$ because of its relatively large longitudinal sound velocity, which together with the smaller $|g|^2$ lead to its weaker LA scattering.

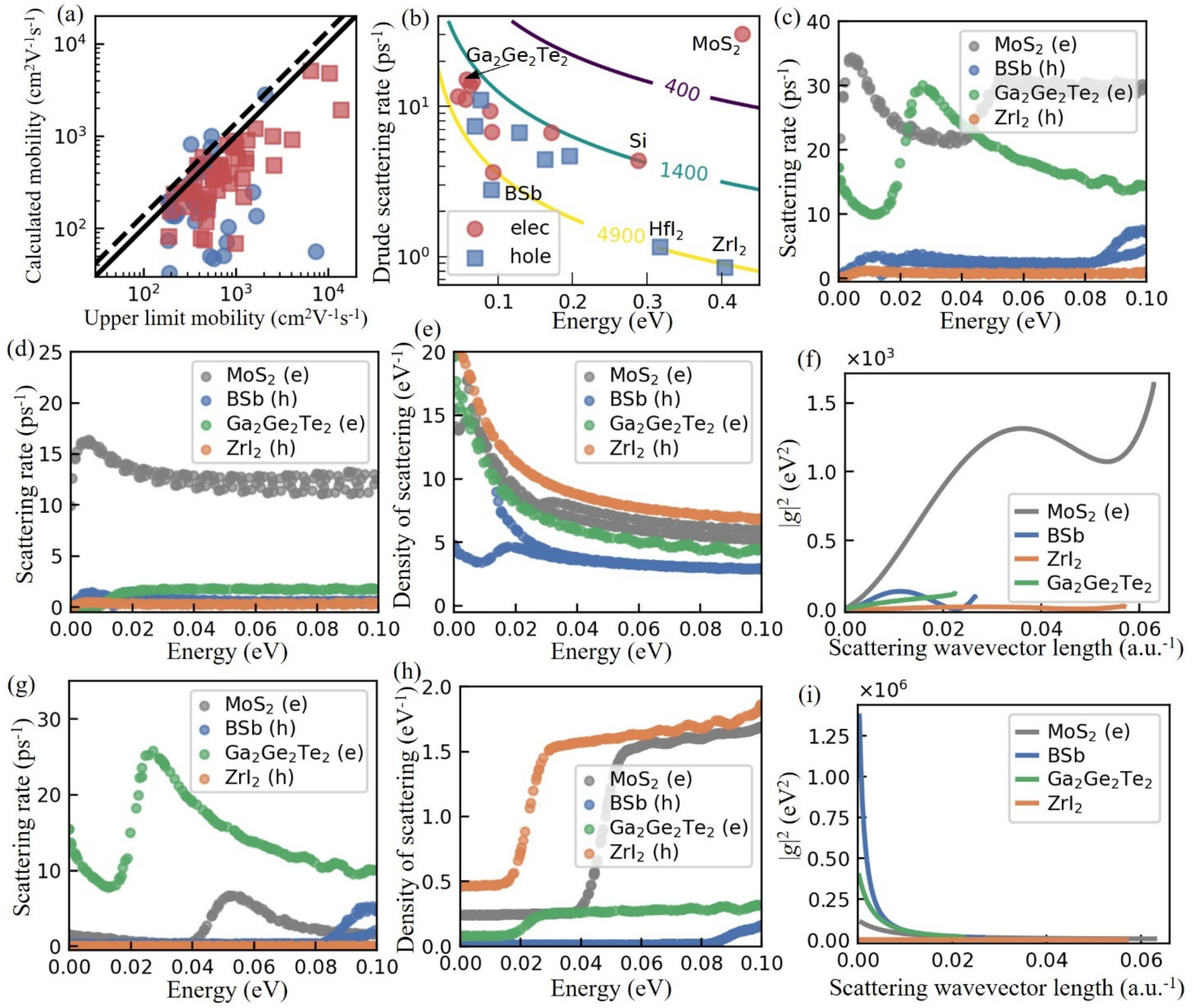

**Figure 8** (a) First-principles calculated mobility vs. upper limited mobility in Ref. 16. (b) Drude scattering rate and Drude effective mass for high-mobility (> 1400 cm$^2$ V$^{-1}$ s$^{-1}$) 2D semiconductors from high-throughput calculation. The lines show the iso-mobility contours. (c) Scattering rates of three representative high-mobility materials. For comparison, $MoS_2$ and Si data are also shown. (d)-(i) Phonon-mode-resolved scattering rates (d and g), density of scatterings (e and h), and representative (squared) EPC $g$ matrix (f and i) for BSb hole, $ZrI_2$ hole, and $Ga_2Ge_2Te_2$ electron, in comparison with $MoS_2$ electron. For above materials, the initial states are selected to be 30 meV above the CBM along the Γ-M direction, and the final states are located at the isoenergy circle of the same valley.

The above analysis shows that the scattering rates can be correlated with basic physical features via $D^S$ and $|g|^2$. For example, the large $R_{B/P}$ leads to large LO $|g|^2$ and large LO scattering for $Ga_2Ge_2Te_2$ and a negligible $R_{B/P}$ in $ZrI_2$ leads to small LO $|g|^2$ and small LO scattering. In spite of BSb having the largest $R_{B/P}$, it has highest LO frequency ($\omega_{LO}$), leading to small $D^S$ and thus weak LO scattering. For LA mode, $Ga_2Ge_2Te_2$, BSb and $ZrI_2$ all have weak LA $|g|^2$, resulting in their weak LA scatterings compared to $MoS_2$. BSb has a highest sound velocity ($v_{LA}$) and hence lowest $D^S$ in 4 materials, which also contributes to its weak LA scattering. The direct correlation between mobility vs. these physical features $M$, $R_{B/P}$, $v_{LA}$ and $\omega_{LO}$ are shown in Figure 9. Additionally, a new feature carrier-lattice distance $d_{c-l}$ is proposed[16] and correlated with mobility in Figure 9c, which is defined as:

$$d_{c\text{-}l}(\mathrm{CBM/VBM}) = \int_{uc} d\mathbf{r}\,|\psi_{\mathrm{CBM/VBM}}(\mathbf{r})|^2 \min_\alpha\{|\mathbf{r}-\mathbf{R}_\alpha|\} \tag{51}$$

where the CBM/VBM indicates the electronic state at conduction band minimum or valence band maximum, the $\psi$ is the corresponding wavefunction, $\mathbf{R}_\alpha$ is the position of nucleus $\alpha$, and uc denotes the unit cell. This new feature $d_{c-l}$ quantifies the distance between the carrier (represented by the CBM/VBM) and the lattice, as illustrated in Fig. 9f. Since the perturbation induced by lattice displacement is generally weaker in the region farther from the nuclei, it is intuitive to expect that a larger $d_{c-l}$ will result in a smaller $|g|^2$. Indeed, as shown in Figure 9e, the high-mobility 2D semiconductors all have a $d_{c-l} > 1.11$ Å.

Therefore, combining Ref. 15 and Ref. 16, the general genome of high-mobility 2D semiconductors include small $M$, small $R_{B/P}$, large $v_{LA}$, large $\omega_{LO}$, small $d_{c-l}$ and anisotropic band edge. The former 5 features are quantified in Ref. 16: high-mobility 2D semiconductor ($\mu$>1400 cm$^2$V$^{-1}$s$^{-1}$) have $M < 0.474$, $n_{LO}\,R^2_{B/P} < 0.066$ Å$^{-1}$ ($n_{LO}$ is the number of LO phonons), $d_{c-l} > 1.11$ Å and $\omega_{LO} > 15$ meV and the anisotropic band edge feature can be quantified by the velocity ($v_{max} > 1.7v_{min}$) as suggested in Ref. 15. The combination of these genome could create more high-mobility 2D semiconductor, particularly including those are not included in Ref. 15 and Ref. 16.

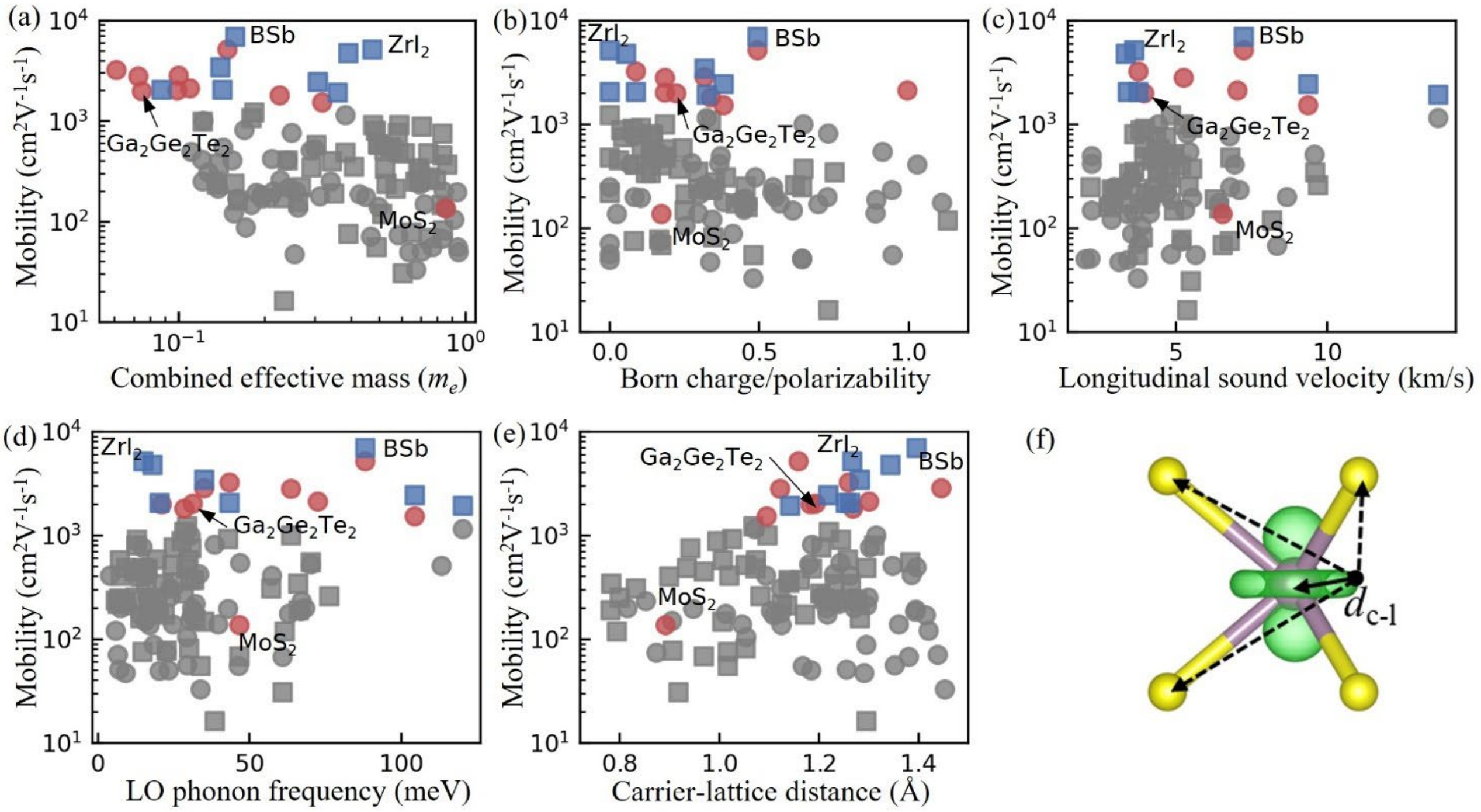

**Figure 9** Phonon-limited mobility vs. various basic physical features: combined effective mass (a), ratio of Born charge to in-plane polarizability (b), longitudinal sound velocity (c), and LO phonon frequency (d), and carrier-lattice distance (e; see text for definition and f for illustration) from Ref. 16.

The high-mobility (> 1400 $cm^2$ $V^{-1}$ $s^{-1}$) 2D semiconductors are highlighted by red (for electron) and blue (for hole).

## 7. Computational approaches in progress

Although the computational approaches of carrier mobility have developed rapidly, they are still not perfect. In Figure 10, we compared the experimentally reported mobility and computational phonon-limited mobility for several well-studied 2D semiconductors, including $MoS_2$ electron[80], $WSe_2$ hole[14], $MoS_2$ hole[131] and $WS_2$ electron[52]. It can be seen that there is still discrepancy between the experimental mobility and the computational one. For example, the experimental mobility of $MoS_2$ hole and $WS_2$ electron is lower than predicted phonon-limited mobility, which could be attributed to other extrinsic scattering. However, for $MoS_2$ electron and $WSe_2$ hole, the experimental mobility is larger than the phonon-limited mobility, implying the underestimation of phonon-limited mobility. Indeed, in realistic devices, there are more effects which might differ the mobility but not well incorporated in current computational regime, including free carrier screening for e-ph/e-d scattering, environmental dielectric screening, substrate effect on band structure, … etc. In this section, we review the recent development of computational approaches for first-principles mobility, which aim to accurately simulate the carrier transport in realistic devices.

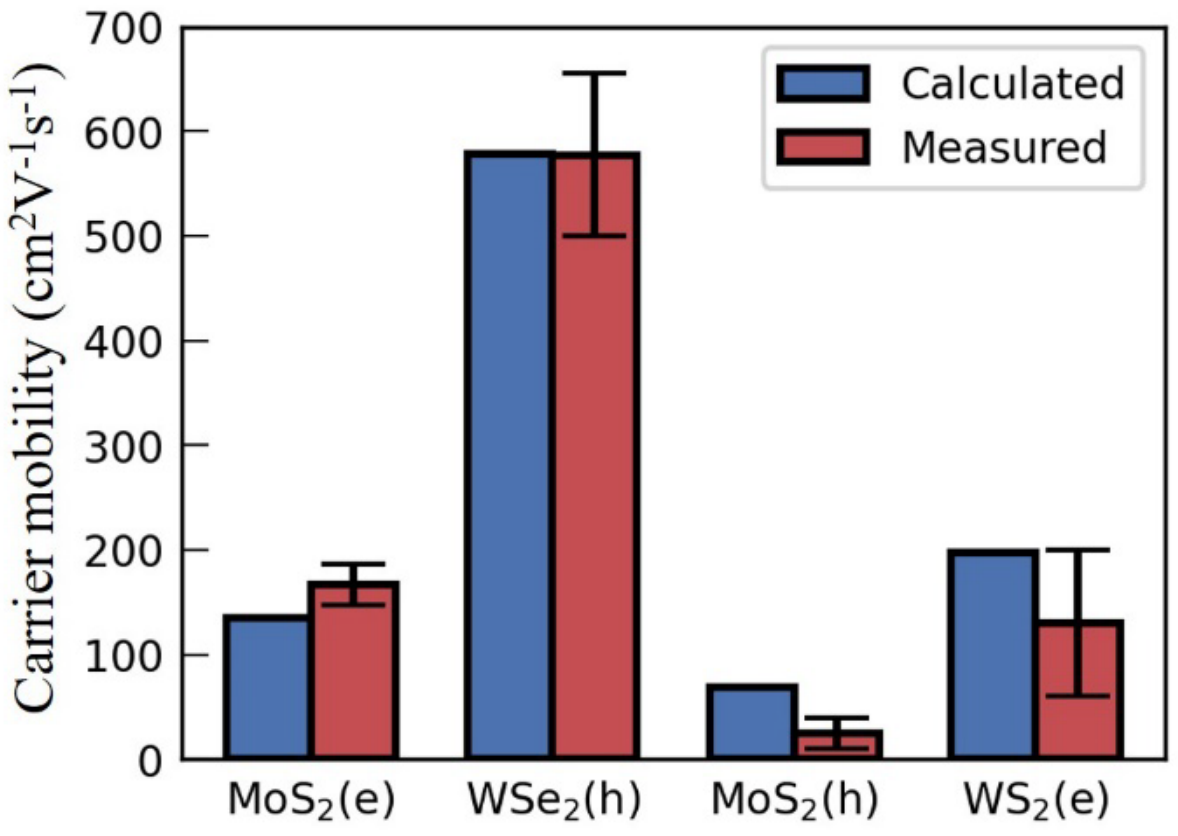

**Figure 10** Room temperature mobility comparison between computations (blue) and experimental works (red) for several well-studied 2D semiconductor monolayers, including $MoS_2$ (electron)[80], $WSe_2$ (hole)[14], $MoS_2$ (hole)[131] and $WS_2$ (electron)[52]. The error bar indicates the highest and lowest mobility reported in the literature.

### 7.1. Free carrier screening in EPC

Due to the atomical thickness, the carrier concentration $n_c$ in 2D semiconductors can be efficiently tuned by electrostatic gate, especially as channels in transistors. The study on free carrier screening effect on EPC is of significance because: (1) the $n_c$ in the channel of transistor varies from intrinsic limit at off-state to as high as $2\times10^{13}$ cm$^{-2}$ at on-state. (2) the free carrier doping would effectively suppress the long-range scattering like Fröhlich scattering, which is the dominant scattering in many polar 2D semiconductors[67], and thus is able to strongly affect the phonon-limited mobility in 2D semiconductors. As mentioned in Section 6, the phonon-limited hole mobility in intrinsic $WSe_2$ is calculated as 578 cm$^2$V$^{-1}$s$^{-1}$ while is 1962 cm$^2$V$^{-1}$s$^{-1}$ when high $n_c$ ($10^{13}$ cm$^{-2}$) is considered[15,16].

There are several computational approaches to include the free carrier screening in e-ph scattering. Here we review their efficiency and accuracy, and try to classify them into different levels of approximation. Starting from Fermi's golden rule, the EPC $g$ matrix in a doped semiconductor is:

$$g_{n'n\nu}(\mathbf{k},\mathbf{q},n_c,T) = \langle \psi_{n'\mathbf{k}+\mathbf{q}}(n_c,T)\,|\,V_{\nu\mathbf{q}}(n_c,T)\,|\,\psi_{n\mathbf{k}}(n_c,T)\rangle, \tag{52}$$

where we explicitly write down the carrier concentration $n_c$ and temperature $T$ dependence in wavefunction $\psi$ and perturbation potential $V$. The $V$ can be decomposed into long-range perturbation potential $V^{LR}$ and short-range perturbation potential $V^{SR}$:

$$V_{\nu\mathbf{q}}(n_c,T) = V_{\nu\mathbf{q}}^{LR}(n_c,T) + V_{\nu\mathbf{q}}^{SR}(n_c,T), \tag{53}$$

by decomposing the bare Coulomb kernel $v$ into long-range part $v^{LR}$ and short-range part $v^{SR}$:

$$v = v^{LR} + v^{SR}. \tag{54}$$

The common choice of $v^{SR}$ is:

$$v_{\mathbf{G}}^{SR}(\mathbf{q}) = (1-\delta_{\mathbf{G}0})\frac{4\pi}{|\mathbf{q}+\mathbf{G}|^2}, \tag{55}$$

in 3D material where **G** is reciprocal-space Bravais lattice and **q** is the wave vector in the Brillouin zone, and is:

$$v_{n,\mathbf{G}_{\|}}^{SR}(\mathbf{q}) = \frac{4\pi}{|\mathbf{q}+\mathbf{G}_{\|}|^2 + G_n^2}[1-(-1)^n], \tag{56}$$

for 2D material in supercell context, where $\mathbf{G}_{\|}$ is reciprocal lattice in the material plane and $G_n = n\pi/L$ along out-of-plane direction and $L$ is range separation length ($2L$ is supercell length). The selection of $v^{SR}$ in Eq. 55 and Eq. 56 facilities the explicit approximation expression of $V^{LR}$ (and $g^{LR}$) for intrinsic 3D[38,43] and 2D semiconductors[25,45] while leaves the $V^{SR}$ ($g^{SR}$) to be determined by first-principles calculations. The free carrier screening effect can be treated in different approaches for $V^{LR}$ ($g^{LR}$) and $V^{SR}$ ($g^{SR}$). In the following, we classify the different computational approaches of free carrier screening effect on EPC into three types based on what approximations are used in $\psi$, $V^{LR}$ and $V^{SR}$.

(1) Full free carrier screening effect

The free carrier screening effects on $\psi$, $V^{\mathrm{LR}}$, $V^{\mathrm{SR}}$ and thus EPC $g$ matrix can be fully considered in a gated DFPT calculation, where additional free carriers in 2D semiconductors are explicitly simulated at first-principles level. The compensating charge (which is required by the convergence of computation) is confined at the virtual gate, below and/or above the 2D layer. It is found that larger $n_{\mathrm{c}}$ effectively suppresses Fröhlich scattering from LO phonons in GaSe[132] while counter-intuitively leads to larger EPC $g$ matrix in multivalley materials like transition metal dichalcogenides (TMDs)[133,134]. The rich phenomena indicate that the free carrier screening effects on EPC is non-trivial. This approach explicitly considers the free carriers, brings least approximations but also heavy computational loads. This is because: (1) Finer **k** grid is required for sampling smaller Fermi surface at lower $n_{\mathrm{c}}$, in both DFT and DFPT calculations. Usually large $n_{\mathrm{c}}$ ($\geq 10^{13}$ cm$^{-2}$) is adapted to maintain a reasonable computational cost. (2) Individual first-principles calculations are required for simulating semiconductors at different $n_{\mathrm{c}}$. (3) The interpolation of EPC $g$ matrix (or interpolation of $V$) is not available, leading to more DFPT calculations involved. Recently, the non-uniform **k** grid sampling is applied to gated DFPT calculations, with more **k** points sampled around the Fermi surface, which is hopeful to speed up the calculations at low-doping regime while keep accuracy[134].

(2) First-order approximation: exact $V^{\mathrm{LR}}$

In a recent study on doped 3D semiconductor SiC[135], the EPC is approximated by neglecting the carrier and temperature dependence for $\psi$, $V^{\mathrm{SR}}$:

$$\begin{aligned} &|\psi_n(n_{\mathrm{c}},T)\rangle \approx |\psi_n(n_{\mathrm{c}}=0,T=0)\rangle, \\ &V^{\mathrm{SR}}(n_{\mathrm{c}},T) \approx V^{\mathrm{SR}}(n_{\mathrm{c}}=0,T=0), \end{aligned} \tag{57}$$

while keeping the $V^{\mathrm{LR}}$ to be exact: $V^{\mathrm{LR}}=V^{\mathrm{LR}}(n_{\mathrm{c}}, T)$. In such a way, the $V^{\mathrm{SR}}$, which is the major cost of computations, can be extracted from single calculation on intrinsic semiconductor and repeatedly used for mobility calculations at different $n_{\mathrm{c}}$. And the easy-to-evaluate $V^{\mathrm{LR}}$ gives free-carrier-screened $g^{\mathrm{LR}}$ at various $n_{\mathrm{c}}$, which can be added back to $g^{\mathrm{SR}}$ (calculated from $V^{\mathrm{SR}}$ in intrinsic system) to recover full $g$ in doped semiconductor. The similar procedure can be applied to 2D doped semiconductor, as shown in Ref. 136. The Ref. 136 found that free carriers significantly change the quadrupolar components in $V^{\mathrm{LR}}$ via "local-field" components of response function while have little effects on dipoles. With the accurate $V^{\mathrm{LR}}$ (determined by screened effective mass $Z$ in Ref. 136) obtained from first-principles (DFPT) calculations, the EPC $g$ matrix and thus relaxation time with free carrier screening can be calculated.

(3) Zero-order approximation: neglecting "local-field" components

The calculations on EPC $g$ matrix with free carrier doping can be further simplified by approximating $V^{\mathrm{LR}}$ via an explicit expression. In addition to assumptions on $\psi$, $V^{\mathrm{SR}}$ (Eq. 57), the $V^{\mathrm{LR}}$ can be further approximated by[25]:

$$V_{\nu\mathbf{q}}^{\mathrm{LR}}(n_{\mathrm{c}},T) \approx \frac{\epsilon_M(\mathbf{q},n_{\mathrm{c}}=0,T=0)}{\epsilon_M(\mathbf{q},n_{\mathrm{c}},T)} V_{\nu\mathbf{q}}^{\mathrm{LR}}(n_{\mathrm{c}}=0,T=0), \tag{58}$$

where $\epsilon_M$ is macroscopic dielectric function in 3D or 2D ($\epsilon^{\|}$ in Eq. 29 for 2D). As $V^{\mathrm{LR}}(n_\mathrm{c}=0,T=0)$ can be evaluated by explicit expression (see Eq. 29 for 2D and Refs. 38,43 for 3D) and $\epsilon_M$ can be easily calculated under random phase approximation (RPA), the $V^{\mathrm{LR}}(n_\mathrm{c},T)$ and thus $g^{\mathrm{LR}}(n_\mathrm{c},T)$ can be directly calculated by Eq. 58 and Eq. 29 for doped 2D semiconductors with specific $n_\mathrm{c}$ and $T$. The above approximation in Eq. 58 is referred to "local-field" approximation since it can be related with a "local-field" approximation in free-carrier-induced independent-particle polarizability $\Delta\chi^0$:

$$\chi^0_{\mathbf{GG}'}(\mathbf{q},n_\mathrm{c},T) \approx \chi^0_{\mathbf{GG}'}(\mathbf{q},n_\mathrm{c}=0,T=0) + \delta_{\mathbf{G}0}\delta_{\mathbf{G}'0}\Delta\chi^0_{3\mathrm{D}}(\mathbf{q},n_\mathrm{c},T), \tag{59}$$

for 3D or

$$\chi^0_{nn',\mathbf{G}_\|\mathbf{G}'_\|}(\mathbf{q},n_\mathrm{c},T) \approx \chi^0_{nn',\mathbf{G}_\|\mathbf{G}'_\|}(\mathbf{q},n_\mathrm{c}=0,T=0) + \delta_{n0}\delta_{n'0}\delta_{\mathbf{G}0}\delta_{\mathbf{G}'0}\Delta\chi^0_{2\mathrm{D}}(\mathbf{q},n_\mathrm{c},T), \tag{60}$$

for 2D. The Eqs. 59 and 60 imply that the free carrier effect on $\chi^0$ is wrapped in an additive $\Delta\chi^0$, which can be modeled by free electron gas (see Appendix C for justifications and details). Write Eqs. 59 and 60 in a uniform form, we have:

$$\chi^0(\mathbf{q},n_\mathrm{c},T) = \chi^0(\mathbf{q},n_\mathrm{c}=0,T=0) + \Delta\chi^0(\mathbf{q},n_\mathrm{c},T). \tag{61}$$

Combining the concrete definitions of $v^{\mathrm{SR}}$ in Eqs. 55 (for 3D) and 56 (for 2D) and the expressions of $\Delta\chi^0$ in Eqs. 59 (for 3D) and 60 (for 2D), it can be seen $v^{\mathrm{SR}}\,\Delta\chi^0=0$, which means the $v^{\mathrm{SR}}\,\chi^0$ is invariant to carrier concentration $n_\mathrm{c}$ and temperature $T$:

$$v^{\mathrm{SR}}(\mathbf{q})\chi^0(\mathbf{q},n_\mathrm{c},T) = v^{\mathrm{SR}}(\mathbf{q})\chi^0(\mathbf{q},n_\mathrm{c}=0,T=0). \tag{62}$$

Eq. 62 brings many desirable properties on $V^{\mathrm{SR}}$ and $V^{\mathrm{LR}}$. As shown in Appendix B, the $V^{\mathrm{SR}}$ is determined by $(\epsilon^{\mathrm{SR}})^{-1}$ and $\epsilon^{\mathrm{SR}}=1-v^{\mathrm{SR}}\chi^0$. Therefore, we have $\epsilon^{\mathrm{SR}}(n_\mathrm{c},T)=\epsilon^{\mathrm{SR}}(n_\mathrm{c}=0,T=0)$ and consequently $V^{\mathrm{SR}}(n_\mathrm{c},T)=V^{\mathrm{SR}}(n_\mathrm{c}=0,T=0)$, which means the $\epsilon^{\mathrm{SR}}$ and $V^{\mathrm{SR}}$ are both independent of $n_\mathrm{c}$ and $T$. The free-carrier independent $\epsilon^{\mathrm{SR}}$ also occurs in $V^{\mathrm{LR}}$ and the only $n_\mathrm{c}$-dependent term in $V^{\mathrm{LR}}$ is the screened long-range interaction $W^{\mathrm{LR}}$ (see Eq. 78 for definition). The $W^{\mathrm{LR}}$ is proportional to $1/\epsilon_\mathrm{M}(n_\mathrm{c},T)$ ($1/\epsilon^{\|}(n_\mathrm{c},T)$ for 2D; see Appendix C for details):

$$W^{\mathrm{LR}}(\mathbf{q},n_\mathrm{c},T) = \frac{\epsilon_M(\mathbf{q},n_\mathrm{c}=0,T=0)}{\epsilon_M(\mathbf{q},n_\mathrm{c},T)} W^{\mathrm{LR}}(\mathbf{q},n_\mathrm{c}=0,T=0), \tag{63}$$

and thus for $V^{\mathrm{LR}}$ we have similar identity shown in Eq. 58. In conclusion, the "local-field" approximation on free-carrier-induced $\Delta\chi^0$ (Eqs. 59 and 60) leads to $V^{\mathrm{SR}}(n_\mathrm{c},T) = V^{\mathrm{SR}}(n_\mathrm{c}=0,T=0)$ and explicit calculation equation of $V^{\mathrm{LR}}(n_\mathrm{c},T)$ (Eq. 58), and thus facilitates the evaluation of overall perturbation potential $V(n_\mathrm{c},T)$. It should be noted that the "local-field" approximation is also adopted in Ref. 135 in a different format: $\bar{Z}(\mathbf{q},n_\mathrm{c},T) \approx \bar{Z}(\mathbf{q},n_\mathrm{c}=0,T=0)$.

Another advantage of the "local-field" approximation depicted above is that it can be efficiently incorporated with interpolation approaches. The $g^{\mathrm{SR}}$ can be evaluated and interpolated from intrinsic 2D semiconductors as the $V^{\mathrm{SR}}$ is independent of $n_\mathrm{c}$. While the $g^{\mathrm{LR}}(n_\mathrm{c},T)$ can be calculated from $g^{\mathrm{LR}}(n_\mathrm{c}=0, T=0)$ and $\epsilon_M(n_\mathrm{c},\mathrm{T})$ by explicit expressions (see Eq. 29 and 58) at different $n_\mathrm{c}$. The EPC $g$ matrix is recovered by combining $g^{\mathrm{SR}}$ and $g^{\mathrm{LR}}$, which then gives relaxation time and carrier mobility. In spite of most approximations, the "local-field" approximation captures the most significant free carrier screening effect in Fröhlich and piezoelectric scattering and the correct trend of mobility change with increasing $n_c$. The gated DFPT calculations considering full free carrier screening effect give 473 $cm^2V^{-1}s^{-1}$ electron mobility in InSe and 1962 $cm^2V^{-1}s^{-1}$ hole mobility in $WSe_2$ both at $n_c=10^{13}$ $cm^{-2}$. In comparison, the "local-field" approximation gives 444 $cm^2V^{-1}s^{-1}$ and 1556 $cm^2V^{-1}s^{-1}$, which are about 10-20% lower than that from gated DFPT approach. The mobility difference can be attributed to screening effect in local fields not included in the "local-field" approach. Despite the differences, the large mobility increase is captured by the "local-field" approach, considering the intrinsic mobilities of InSe and $WSe_2$ are 117 and 578 $cm^2V^{-1}s^{-1}$.

## 7.2. Charged defect and free carrier screening

In semiconductors with shallow defects (i.e. defect level is close to CBM or VBM) at room temperature, the carrier in defect band would be thermalized and charge the defects, leading to charged defect with strong Coulomb potential. The scattering between the carrier and the charged defect (denoted as 'e-cd' scattering hereafter) is more complicated than the neutral defect (e-d) scattering introduced in Section 3.2. First, simulating charged defect in a supercell geometry, especially in 2D semiconductor, is still an open challenge. In contrast to simulation of neutral defect, the extra charge localized around charged defect has fictitious images due to repetition of the supercell and leads to divergence of the Coulomb energy. The commonly used solution for 3D semiconductor like jellium compensating charge would lead to artificial bound states in the vacuum for 2D charged defect, resulting in erroneous formation energy and band structures[137]. Secondly, the extrinsic free carrier screening plays an essential role in e-cd scattering, as the charged defect has a long-tail Coulomb potential which can be efficiently screened by carriers. How these free carriers will interact with charged defect and whether via a localized screening (forming charge trapped states or "mobility edge" in literatures)[138-140] or delocalized screening regime are still elusive.

Therefore, a fully accurate calculation of e-cd scattering with free carrier screening effect is still not available. The difficulty can be circumvented by substituting the charged defect perturbation potential with model potentials. In Ref. 141, the perturbation potential in 3D semiconductor is approximated by a point charge Coulomb potential screened by a model dielectric function which takes free carriers into account via Lindhard dielectric model. In Ref. [defect xx], a more delicate dielectric function, including intrinsic screening from GW method and free carrier screening from Lindhard model, is used for charged S/Se vacancy in 2D TMDs. A better approximation involves separating the $V^{LR}$ and $V^{SR}$ to better deal with free carrier screening. In Ref. 142, charged defect in 3D Si is simulated in supercell without extrinsic free carriers, then a long-range Coulomb potential (denoted as $V^{LR}$ here) is extracted and screened by additional free carriers while the remaining short-range potential $V^{SR}$ is kept intact. Then the screened $V^{LR}$ and original $V^{SR}$ are combined to give full perturbation potential V and thus EDI $M$ matrix. The procedure in Ref. 142 resembles the zero-order approximation for free carrier screening in EPC (see section 7.1). However, the charged defect simulation in this approach relies on jellium compensating charge which, as mentioned above, is not reliable for 2D semiconductor. So its generalization to charged defect in doped 2D semiconductor is still unavailable.

### 7.3. Environmental dielectric screening

Besides the free carrier screening, the dielectric environment also provides an effective approach to screen long-range e-p and e-cd scattering. As shown in Ref. 111, the InSe phonon-limited mobility can be improved from 120 to 525 $cm^2V^{-1}s^{-1}$ by tuning environmental dielectric constant and suppressing the Fröhlich scattering in InSe. In Ref. 132, a highly doped graphene is proposed as a long-range EPC screening layer when separated by a BN monolayer from channel layer GaSe. In such heterostructure Graphene/BN/GaSe, the monolayer GaSe phonon-limited mobility is predicted to improve from 174 to 500-600 $cm^2V^{-1}s^{-1}$ even with low $n_c$ ($10^{11}$ $cm^{-2}$). A more prominent example of environmental dielectric screening is for e-cd scattering. In experiments, high dielectric constant insulators are usually applied to suppress the charged defect scattering and improve carrier mobility in 2D transistor[97,143-145]. However, the theoretical studies usually apply model charged defect potential [146,147]. A first-principles calculation of charged defect scattering accurately considering environmental dielectric screening is highly in demand, especially for transistor applications.

### 7.4. Substrate effect on band structure

In addition to screening, another factor that leads to the discrepancy between experimentally reported carrier mobility and those from computations is the substrate-induced band structure change. In calculations, the 2D semiconductor is usually simulated as a free-standing monolayer. However, in realistic devices, the 2D semiconductor is usually placed on substrate or encapsulated by other vdW 2D layers, which introduces variation of band structures. In multivalley semiconductors like TMDs, the small band structure variation, especially the energy separation between K and Q valleys ($\Delta E_{KQ}$) would significantly change the carrier mobility. In Ref. 148, the 0.4% increase of layer thickness of $WS_2$ would result in 36 meV decrease of $\Delta E_{KQ}$ and thus 50% decrease of electron mobility. A more thorough study[134] found that small $\Delta E_{KQ}$ has a detrimental effect on phonon-limited mobility for all TMDs, not only due to it facilitating intervalley scattering but also strengthening the EPC of intervalley scattering.

However, the calculation of accurate quasi-particle band structure is still a challenge. Ref. 149 shows that the calculated $\Delta E_{KQ}$ depends on the selection of pseudopotential and exchange-correlation functional. Higher level of computational methods like $G_0W_0$ calculation with full geometry relaxation[150] or self-consistent GW calculation[151,152] are required to reproduce the positive $\Delta E_{KQ}$ (which means direct bandgap) found in experiments. Moreover, it is found that existence of substrate ($SiO_2$ and hBN) will change the screened Coulomb interaction in 2D $MoS_2$ and thus the band structure, especially leading to modification of band gap[153,154]. The $\Delta E_{KQ}$ with different substrates is predicted to change nearly 100 meV[150] for $MoS_2$ while reported to be small[154] in experiment on $WS_2$. In general, the accurate transport calculation in TMDs with fully considering the correct band structure is still not available, which might account for the difference between experiments and computations.

### Appendix A: Explicit definition of EPC strength *D*

From the definitions of EPC $g$ matrix and perturbation potential $V$ in Eqs. 25 and 26, the $g$ can be written as:

$$g_{n'n\nu}(\mathbf{k},\mathbf{q}) = l_{\nu\mathbf{q}} \langle \psi_{n'\mathbf{k}+\mathbf{q}} | e^{i\mathbf{q}\cdot\mathbf{r}} \sum_{\kappa\alpha} \sqrt{\frac{M_0}{M_\kappa}} e_{\kappa\alpha,\nu}(\mathbf{q}) \frac{\partial V_{\mathrm{KS}}^{(\mathrm{ph})}}{\partial u_{\kappa\alpha}(\mathbf{q})} | \psi_{n\mathbf{k}} \rangle. \tag{64}$$

Recall that the EPC strength $D_{n'n\nu}(\mathbf{k},\mathbf{q}) = g_{n'n\nu}(\mathbf{k},\mathbf{q}) / l_{\nu\mathbf{q}}$, therefore the $D$ is:

$$D_{n'n\nu}(\mathbf{k},\mathbf{q}) = \langle \psi_{n'\mathbf{k}+\mathbf{q}} | e^{i\mathbf{q}\cdot\mathbf{r}} \sum_{\kappa\alpha} \sqrt{\frac{M_0}{M_\kappa}} e_{\kappa\alpha,\nu}(\mathbf{q}) \frac{\partial V_{\mathrm{KS}}^{(\mathrm{ph})}}{\partial u_{\kappa\alpha}(\mathbf{q})} | \psi_{n\mathbf{k}} \rangle. \tag{65}$$

The $\frac{\partial V_{\mathrm{KS}}^{(\mathrm{ph})}}{\partial u_{\kappa\alpha}(\mathbf{q})}$ is the derivative of Kohn-Sham (KS) potential due to periodic atom displacement, which is defined as:

$$\frac{\partial V_{\mathrm{KS}}^{(\mathrm{ph})}}{\partial u_{\kappa\alpha}(\mathbf{q})} = \sum_{p} e^{-i\mathbf{q}\cdot(\mathbf{r}-\mathbf{R}_p)} \frac{\partial V_{\mathrm{KS}}^{(\mathrm{ph})}}{\partial \tau_{p\kappa\alpha}}, \tag{66}$$

where $\frac{\partial V_{\mathrm{KS}}^{(\mathrm{ph})}}{\partial \tau_{p\kappa\alpha}}$ is the KS potential perturbation due to atom $\kappa$ in $p$-th unit cell moving in $\alpha$ direction and $\mathbf{R}_p$ is the spatial coordinates of $p$-th unit cell. Combining Eqs. 63 and 64, the $D$ can be written as:

$$D_{n'n\nu}(\mathbf{k},\mathbf{q}) = \langle \psi_{n'\mathbf{k}+\mathbf{q}} | \sum_{p\kappa\alpha} \sqrt{\frac{M_0}{M_\kappa}} e_{\kappa\alpha,\nu}(\mathbf{q}) e^{i\mathbf{q}\cdot\mathbf{R}_p} \frac{\partial V_{\mathrm{KS}}^{(\mathrm{ph})}}{\partial \tau_{p\kappa\alpha}} | \psi_{n\mathbf{k}} \rangle \tag{67}$$

which is exactly Eq. 33.

**Appendix B: Range separation in EPC perturbation potential $V$**

Before considering the range-separation of $V$, first we briefly review the case without range-separation. In linear response theory, the screened potential $V$ in materials under static external potential $U^{\mathrm{ext}}$ is:

$$V(\mathbf{r}, n_{\mathrm{c}}, T) = \int \epsilon^{-1}(\mathbf{r}, \mathbf{r}', n_{\mathrm{c}}, T) U^{\mathrm{ext}}(\mathbf{r}') d\mathbf{r}', \tag{68}$$

where $\epsilon^{-1}$ is inverse dielectric function with explicit dependence to carrier concentration $n_{\mathrm{c}}$ and temperature $T$. Written as operators, we have:

$$V = \epsilon^{-1} U^{\mathrm{ext}}. \tag{69}$$

The $\epsilon^{-1}$ is directly related to the reducible polarizability $\chi$ by:

$$\epsilon^{-1} = 1 + v\chi, \tag{70}$$

where $v$ is the Coulomb kernel. The $\epsilon^{-1}$ describes the screened potential $V$ under static external potential $U^{\text{ext}}$, and $\chi$ linearly relates the (induced) charge response $\rho^{\text{ind}}$ with the $U^{\text{ext}}$. The $\epsilon^{-1}$ and $\chi$ can be calculated by independent-particle polarizability $\chi^0$ (Eq. 82) by:

$$\epsilon = 1 - v\chi^0, \tag{71}$$

and a Dyson equation:

$$\chi = \chi^0 + \chi^0 v\chi. \tag{72}$$

The Eqs. 68, 69 and 70 are key equations for linear-response approach for EPC perturbation potential $V$, where $U^{\text{ext}}$ can be interpreted as the bare potential induced by nuclei movement in phonons.

Ref. 46 shows that if the bare Coulomb kernel $v$ is separated into a short-range (SR) and a long-range (LR) part:

$$v = v^{\text{LR}} + v^{\text{SR}} \tag{73}$$

then the perturbation potential $V$ can be correspondingly separated into two parts:

$$V = V^{\text{LR}} + V^{\text{SR}}. \tag{74}$$

The $V^{\text{SR}}$ is:

$$V^{\text{SR}} = (\epsilon^{\text{SR}})^{-1} v^{\text{SR}} \rho^{\text{ext}} \tag{75}$$

where $\rho^{\text{ext}}$ is the "external" charge perturbation potential determined by $U^{\text{ext}}$, and $\epsilon^{\text{SR}}$ is the short-range dielectric function:

$$\begin{aligned} \rho^{\text{ext}} &= v^{-1} U^{\text{ext}}, \\ \epsilon^{\text{SR}} &= 1 - v^{\text{SR}} \chi^0. \end{aligned} \tag{76}$$

Since the $V^{\text{SR}}$ has a similar expression to $V = \epsilon^{-1} v \rho^{\text{ext}}$ except only partial of $v$ (i.e. $v^{\text{SR}}$) is considered in $V^{\text{SR}}$, the $V^{\text{SR}}$ can be interpreted as the screened perturbation potential when electron interaction in system is $v^{\text{SR}}$. The remaining long-range perturbation potential $V^{\text{LR}} = V - V^{\text{SR}}$ can be proved to be (see Appendix A in Ref. 46):

$$V^{\text{LR}} = (\epsilon^{\text{SR}})^{-1} W^{\text{LR}} [(\epsilon^{\text{SR}})^{-1}]^{\dagger} \rho^{\text{ext}}. \tag{77}$$

where $W^{\text{LR}}$ is long-range screened Coulomb interaction:

$$W^{\text{LR}} = (1 - v^{\text{LR}} \chi^{\text{SR}})^{-1} v^{\text{LR}} = (1 + v^{\text{LR}} \chi) v^{\text{LR}} \tag{78}$$

In the following, we show that the $W^{\mathrm{LR}}$ can be directly related to macroscopic dielectric function $\epsilon_{\mathrm{M}}(\mathbf{q})$ (in 3D) and $\epsilon^{\|}(\mathbf{q})$ (in 2D). In 3D system, by applying range-separation shown in Eq. 55 (i.e. $v_{\mathbf{G}}^{\mathrm{SR}}(\mathbf{q})=\delta_{\mathbf{G}0}\frac{4\pi}{|\mathbf{q}+\mathbf{G}|^2}$), the $W^{\mathrm{LR}}$ (Eq. 76) is:

$$W_{\mathbf{GG}'}^{\mathrm{LR}}(\mathbf{q},n_{\mathrm{c}},T)=\delta_{\mathbf{G}0}\delta_{\mathbf{G}'0}\frac{4\pi}{|\mathbf{q}|^2}[1+\frac{4\pi}{|\mathbf{q}|^2}\chi_{00}(\mathbf{q},n_{\mathrm{c}},T)]. \tag{79}$$

Recall that $\epsilon^{-1}=1+v\chi$, the $W^{\mathrm{LR}}$ in Eq. 77 thus is:

$$W_{\mathbf{GG}'}^{\mathrm{LR}}(\mathbf{q},n_{\mathrm{c}},T)=\delta_{\mathbf{G}0}\delta_{\mathbf{G}'0}\frac{4\pi}{|\mathbf{q}|^2}\frac{1}{\epsilon_M(\mathbf{q},n_{\mathrm{c}},T)}, \tag{80}$$

where $\epsilon_{\mathrm{M}}$ is the macroscopic dielectric function, defined as the inverse of head term of $\epsilon^{-1}$: $\epsilon_M(\mathbf{q})=1/[\epsilon^{-1}]_{00}$. The $\epsilon_{\mathrm{M}}$ can be obtained by experiments, RPA dielectric function[155] or a model dielectric function[135,142]. Ref. 46 shows that in 2D system, the mirror-even component of $W^{\mathrm{LR}}$ can be approximated by:

$$W_{\mathbf{G}_{\|}\mathbf{G}_{\|}'}^{\mathrm{LR}}(\mathbf{q},n_{\mathrm{c}},T)=\delta_{\mathbf{G}_{\|}\mathbf{G}_{\|}'}\frac{2\pi f(|\mathbf{q}|)}{|\mathbf{q}|}\frac{1}{\epsilon^{\|}(\mathbf{q},n_{\mathrm{c}},T)}, \tag{81}$$

where $\epsilon^{\|}$ is the (macroscopic) in-plane dielectric function for 2D semiconductor[46], $f(q)=1-\tanh(qL/2)$ is range separation function and $L$ is range separation length. From Eqs. 78 and 79, we can see that the $W^{\mathrm{LR}}$ for 2D and 3D are both determined by macroscopic dielectric function.

## Appendix C: Linear response in doped semiconductor and "local-field" approximation

In random-phase approximation, the independent-particle polarizability $\chi^0$ is given by the Adler-Wiser equation:

$$\chi_{\mathbf{GG}'}^{0}(\mathbf{q})=\frac{1}{\Omega}\sum_{nn'\mathbf{k}}\frac{f_n^0(\mathbf{k})-f_{n'}^0(\mathbf{k}+\mathbf{q})}{\varepsilon_{n\mathbf{k}}-\varepsilon_{n'\mathbf{k}+\mathbf{q}}}\langle\psi_{n\mathbf{k}}|e^{-i(\mathbf{q}+\mathbf{G})\cdot\mathbf{r}}|\psi_{n'\mathbf{k}+\mathbf{q}}\rangle\langle\psi_{n'\mathbf{k}+\mathbf{q}}|e^{i(\mathbf{q}+\mathbf{G}')\cdot\mathbf{r}}|\psi_{n\mathbf{k}}\rangle, \tag{82}$$

where $\psi_{n\mathbf{k}}(\psi_{n'\mathbf{k}+\mathbf{q}})$ is electronic state with wavevector $\mathbf{k}(\mathbf{k}+\mathbf{q})$ and band index $n(n')$, $f^0$ is Fermi distribution, $\varepsilon$ is electronic energy. In an intrinsic semiconductor (i.e. carrier concentration $n_{\mathrm{c}}$ and temperature $T$ are assumed to be low enough), the fermi level is located between the band gap and the occupation of state is either 1 (valance band) or 0 (conduction band). Therefore, the $\chi^0$ is totally contributed by interband components, where $n=v$, $n'=c$ ($n$ is from valance band and $n'$ is from conduction band) or $n=c$, $n'=v$ ($n$ from conduction and $n'$ from valence band). The $\chi^0$ is:

$$\chi_{\mathbf{GG}'}^{0}(\mathbf{q},n_{\mathrm{c}}=0,T=0)=\frac{2}{\Omega}\sum_{cv\mathbf{k}}\frac{\langle\psi_{v\mathbf{k}}|e^{-i(\mathbf{q}+\mathbf{G})\cdot\mathbf{r}}|\psi_{c\mathbf{k}+\mathbf{q}}\rangle\langle\psi_{c\mathbf{k}+\mathbf{q}}|e^{i(\mathbf{q}+\mathbf{G}')\cdot\mathbf{r}}|\psi_{v\mathbf{k}}\rangle}{\varepsilon_{v\mathbf{k}}-\varepsilon_{c\mathbf{k}+\mathbf{q}}}, \tag{83}$$

In a doped semiconductor, the Fermi level is shifted closer to CBM or VBM, leading to partial occupation of electrons (holes) in conduction (valence) bands. Intraband components occur in $\chi^0$ as $f_n^0(\mathbf{k}) - f_{n'}^0(\mathbf{k}+\mathbf{q}) \neq 0$ for certain **k** and **q**. By applying "rigid-band model", we assume the doping is small and insufficient to change $\psi_{n\mathbf{k}}$ and $\varepsilon_{n\mathbf{k}}$:

$$\begin{aligned} |\psi_{n\mathbf{k}}(n_c, T)\rangle &= |\psi_{n\mathbf{k}}\rangle, \\ \varepsilon_{n\mathbf{k}}(n_c, T) &= \varepsilon_{n\mathbf{k}}. \end{aligned} \tag{84}$$

So the $\chi^0$ for a doped semiconductor can be written as:

$$\chi^0(\mathbf{q}, n_c, T) = \chi^0(\mathbf{q}, n_c = 0, T = 0) + \Delta\chi^0_{\text{inter}}(\mathbf{q}, n_c, T) + \Delta\chi^0_{\text{intra}}(\mathbf{q}, n_c, T), \tag{85}$$

where $\Delta\chi^0_{\text{inter}}$ indicates the change of interband components due to the change of $n_c$ and $T$ and $\Delta\chi^0_{\text{intra}}$ is the additional intraband components introduced by $n_c$ and $T$. Usually $\Delta\chi^0_{\text{inter}} = 0$ is assumed since the change of $f^0$ occurs in a small region (around CBM or VBM) in whole BZ, which leads to small change of interband $\chi^0$ integrating over whole BZ. However, the $\Delta\chi^0_{\text{inter}}$ leads to a remarkable change of $\chi^0$, especially at zone center, which is due to the vanishing denominator ($\varepsilon_{n\mathbf{k}}-\varepsilon_{n'\mathbf{k+q}}$) as $\mathbf{q}\rightarrow 0$. As discussed in Ref. 29, the $\Delta\chi^0_{\text{inter}}$ would change the **q**-dependence of head and wings term in $\chi^0$. In intrinsic semiconductor, the head of $\chi^0$ scales with $|\mathbf{q}|^2$ and wings are linear with $|\mathbf{q}|$. While in doped semiconductor or metal, the matrix elements of $\chi^0$ have finite limit at $\mathbf{q}=0$.

In "local-field" approximation of free-carrier screening in $\chi^0$, the head term of $\Delta\chi^0_{\text{inter}}$=is only considered and can be further simplified by dielectric models. The $\Delta\chi^0_{\text{inter}}$ (written as $\Delta\chi^0$ hereafter) can be calculated as:

$$\Delta\chi^0_{\text{3D/2D}}(\mathbf{q}, n_c, T) = \frac{1}{\Omega}\sum_{n\mathbf{k}} \frac{f_n^0(\mathbf{k}, n_c, T) - f_n^0(\mathbf{k}+\mathbf{q}, n_c, T)}{\varepsilon_{n\mathbf{k}} - \varepsilon_{n\mathbf{k}+\mathbf{q}}}, \tag{86}$$

where subscript '3D/2D' indicates dimensions of the system. In practical calculations, the macroscopic dielectric function $\epsilon_{\text{M}}(\mathbf{q})$ (and $\epsilon^{||}(\mathbf{q})$) for intrinsic 3D (2D) semiconductor is usually approximated by:

$$\begin{aligned} \epsilon_M(\mathbf{q}, n_c = 0, T = 0) &= \epsilon^{\infty}_{\text{3D}} \\ \epsilon^{||}(\mathbf{q}, n_c = 0, T = 0) &= 1 + 2\pi\alpha_{\text{2D}}(\hat{\mathbf{q}})|\mathbf{q}| \end{aligned} \tag{87}$$

where $\epsilon^{\infty}_{\text{3D}}$ is dielectric constant for 3D semiconductor, $\alpha_{\text{2D}}(\hat{\mathbf{q}}) = \hat{\mathbf{q}}\cdot\boldsymbol{\alpha}_{\text{2D}}\cdot\hat{\mathbf{q}}$ is **q**-direction dependent 2D polarizability and $\boldsymbol{\alpha}_{\text{2D}}$ is 2×2 polarizability tensor for 2D semiconductor. Using the identity $\epsilon = 1 - v\chi^0$ and neglecting the local field (i.e. $\epsilon^{-1}(\mathbf{q}, n_c, T) \approx 1/\epsilon(\mathbf{q}, n_c, T)$), we have approximate equations to evaluate $\epsilon_M(\mathbf{q})$ (and $\epsilon^{\|}(\mathbf{q})$) for doped semiconductor:

$$\begin{aligned} \epsilon_M(\mathbf{q}, n_c, T) &= \epsilon_{3D}^{\infty} - \frac{4\pi}{|\mathbf{q}|^2}\Delta\chi_{3D}^0(\mathbf{q}, n_c, T) \\ \epsilon^{\|}(\mathbf{q}, n_c, T) &= 1 + 2\pi\alpha_{2D}(\hat{\mathbf{q}})|\mathbf{q}| - \frac{2\pi}{|\mathbf{q}|}\Delta\chi_{2D}^0(\mathbf{q}, n_c, T) \end{aligned} \tag{88}$$

The $\Delta\chi_{3D/2D}^0(\mathbf{q}, n_c, T)$ can be calculated by real band structures[135] via Eq. 86 or from parabolic band structure[156] with effective mass $m^*$.

**Appendix D: Mobilities of 2D semiconductors**

| Materials (e for electrons and h for holes) | Measured mobility (in unit of $cm^2V^{-1}s^{-1}$; thickness is indicated in bracket, otherwise monolayer is implied; mobilities are listed in order of published date, from latest to earliest) | Calculated mobility (in unit of $cm^2V^{-1}s^{-1}$; carrier concentration is indicated in bracket in unit of $10^{13}$ $cm^{-2}$, otherwise low carrier concentration limit is implied; mobilities considering dipolar and quadrupolar scattering are boldened; mobilities are listed in order of reported date, from latest to earliest) |
|---|---|---|
| $MoS_2$ (e) | 107±16 (bilayer)[0], 80±20[0], 55[0], 167±20[6], 260 (18.5nm)[4], 125[3], 51[2], 36±22[1], 470 (50nm)[7] | **136** [77], **172**(1)[78], **136**[78], 168[7], 145[10], 183[6], **144**(5)[5], 400[4], 150[3], 130[2], 400[1] |
| $MoS_2$ (h) | 25±15[5], 480 (50nm)[7] | **69** [77], 270[9] |
| $MoSe_2$ (e) | 50[15] | 101[10], 88[6], 25[9] |
| $MoSe_2$ (h) | N.A. | **147** [77], 90[9] |
| $WS_2$ (e) | 130±70[1], 105±45[10], 83±10[6], 54±29[9], 50±10[8] | **197** [77], 215[10], 246[6], **60**(5)[8], 320[9] |
| $WS_2$ (h) | N.A. | **919** [77], 540[9] |
| $WSe_2$ (e) | 14±2[0], 30[13] | 46[10], 53[6], **25**(5)[8], 30[9] |
| $WSe_2$ (h) | 578±78[11], 150[12], 140[14], 180[13] | **578** [77], **1962**(1)[16], 270[9] |
| $In_2Se_2$ (e) | 30[22], 1000 (5.4nm)[21] | **444**(1)[78], **117**[78], **473**(1)[16], 18[6], 118[14], 110[12], 314[11], 120[13] |
| $P_4$ (h) | 984 (10nm)[18], 243±43 (5nm)[17] | **480** [77], **1386**(1)[16] |
| $P_4$ (e) | N.A. | **248** [77], 303[6] |
| AsP (e) | 83 (37nm)[20] | N.A. |
| $Bi_2O_2Se$ (e) | 300±150 (9.8nm)[29] | N.A. |
| $HfSe_2$ (e) | 3 (2nm)[32] | 2[6] |
| $In_2Se_3$ (e) | 400±88 (79nm)[26] | N.A. |
| $MoTe_2$ (e) | 40 (28nm)[16] | 50[10], 41[6] |
| $PdSe_2$ (e) | 158 (12nm)[25] | **104** [77], 82[6] |
| $PtSe_2$ (e) | 210 (11nm)[24] | 39[6] |
| $ReS_2$ (e) | 16 (4.5nm)[31], 1±1[30] | N.A. |
| $ReSe_2$ (e) | 6 (7nm)[31] | N.A. |
| SnSe (e) | 254 (10nm)[23] | N.A. |
| $As_4$ (h) | 48±3[19] | **219** [77] |
| AsP (h) | 79 (37nm)[20] | N.A. |
| $MoTe_2$ (h) | 56 (28nm)[16] | **81** [77] |
| $Nb_2SiTe_4$ (h) | 98 (7.5nm)[27] | N.A. |
| $ReSe_2$ (h) | 5 (7nm)[31] | N.A. |
| Te (h) | 535±125 (16nm)[30] | N.A. |
| $Ag_2Te_2$ (e) | N.A. | **50** [77] |
| $Al_2Ge_2Se_2$ (e) | N.A. | **421** [77] |
| $Al_2Ge_2Te_2$ (e) | N.A. | **2023** [77] |
| $Al_2MgSe$= (e) | N.A. | **217** [77] |
| AlBi (e) | N.A. | **2835** [77] |
| $AlLiTe_2$ (e) | N.A. | **425**(1)[16] |
| $As_2$ (e) | N.A. | **56** [77] |
| $As_4$ (e) | N.A. | **71** [77] |
| $Au_2Te_2$ (e) | N.A. | **51** [77] |
| $B_4C$ (e) | N.A. | 2140[21], 2370[21] |
| BAs (e) | N.A. | **1524** [77] |
| BBi (e) | N.A. | **523** [77] |
| BP (e) | N.A. | **1151** [77] |
| BSb (e) | N.A. | **5167** [77] |
| $Bi_2Se_3$ (e) | N.A. | **345**(1)[16] |
| $Bi_2SeTe_2$ (e) | N.A. | **1100**(1)[16] |
| BiClTe (e) | N.A. | **384**(1)[16] |
| $Cd_2S_2$ (e) | N.A. | **138** [77] |
| $Cd_2Te_2$ (e) | N.A. | **248** [77] |
| $F_2Si_2$ (e) | N.A. | **760** [77] |
| $Ga_2Ge_2Te_2$ (e) | N.A. | **1996** [77] |
| $Ga_2O_2$ (e) | N.A. | **176** [77] |
| $Ga_2S_2$ (e) | N.A. | 7[6] |
| $Ga_2Se_2$ (e) | N.A. | **172** [77], **199** [77], **594**(1)[16], 39[6] |
| $Ga_2Si_2S_2$ (e) | N.A. | **178** [77] |
| $Ga_2Si_2Se_2$ (e) | N.A. | **260** [77] |
| $Ga_2Te_2$ (e) | N.A. | 20[6] |
| GaAs (e) | N.A. | **812** [77] |
| GaN (e) | N.A. | **510** [77] |
| GaP (e) | N.A. | **408** [77] |
| GaSb (e) | N.A. | **1809** [77] |
| $Ge_2H_2$ (e) | N.A. | **2791** [77], 2380[18] |
| $Ge_2Te_2$ (e) | N.A. | **47** [77] |
| $HfS_2$ (e) | N.A. | 1[6] |
| $In_2O_2$ (e) | N.A. | **188** [77] |
| $In_2S_2$ (e) | N.A. | 10[6] |
| $In_2Te_2$ (e) | N.A. | **183** [77], **194** [77], 28[6] |
| InAs (e) | N.A. | **1001** [77] |
| InN (e) | N.A. | **2106** [77] |
| InP (e) | N.A. | **542** [77] |
| $Mo_2Cl_6$ (e) | N.A. | **74** [77] |
| $MoO_2$ (e) | N.A. | **200** [77] |
| $MoSi_2N_4$ (e) | N.A. | 87[17] |
| $Os_2S_4$ (e) | N.A. | **150** [77] |
| $P_2$ (e) | N.A. | 51[6] |
| $Pb_2Te_2$ (e) | N.A. | **170** [77] |
| PbTe (e) | N.A. | **147** [77] |
| $PtS_2$ (e) | N.A. | 24[6] |
| $PtTe_2$ (e) | N.A. | 31[6] |
| $S_2Si$ (e) | N.A. | **67** [77] |
| $Sb_2$ (e) | N.A. | **49** [77] |
| $Sb_2Se_3$ (e) | N.A. | **88** [77] |
| $Sb_2SeTe_2$ (e) | N.A. | **252** [77], **719**(1)[16] |
| $Sb_2Te_3$ (e) | N.A. | **212** [77] |
| $Se_2Si$ (e) | N.A. | **55** [77] |
| $Se_2Si_2$ (e) | N.A. | **33** [77] |
| $Si_2Br_2$ (e) | N.A. | **164** [77] |
| $Si_2Cl_2$ (e) | N.A. | **198** [77] |
| $Si_2H_2$ (e) | N.A. | 24[20] |
| $Sn_2H_2$ (e) | N.A. | **3227** [77] |
| $Sn_2Se_2$ (e) | N.A. | 70[19] |
| $Sn_2Te_2$ (e) | N.A. | **120** [77] |
| $SnS_2$ (e) | N.A. | 50[22], 6[6] |
| $SnSe_2$ (e) | N.A. | 8[6] |
| $Tl_2Te_2$ (e) | N.A. | **413** [77], **494** [77] |
| $WO_2$ (e) | N.A. | **232** [77] |
| $WSi_2N_4$ (e) | N.A. | 119[17] |
| $WTe_2$ (e) | N.A. | **139** [77] |
| $Zn_2Se_2$ (e) | N.A. | **170** [77] |
| $Zn_2Te_2$ (e) | N.A. | **308** [77] |
| $Zr_2Te_2$ (e) | N.A. | **137** [77] |
| $ZrS_2$ (e) | N.A. | 1[6] |
| $ZrSe_2$ (e) | N.A. | 1[6] |
| $Al_2O_2$ (h) | N.A. | **258** [77] |
| AlBi (h) | N.A. | **3446** [77] |
| $As_2$ (h) | N.A. | **1216** [77], 790[15] |
| $Au_2I_2$ (h) | N.A. | **326**(1)[16] |
| $Au_2Te_2$ (h) | N.A. | **248** [77] |
| BAs (h) | N.A. | **2439** [77] |
| BBi (h) | N.A. | **548** [77] |
| BP (h) | N.A. | **1921** [77] |
| BSb (h) | N.A. | **6935** [77] |
| $Cl_2Ga_2Te_2$ (h) | N.A. | **327**(1)[16] |
| $F_2Si_2$ (h) | N.A. | **76** [77] |
| GaAs (h) | N.A. | **16** [77] |
| GaSb (h) | N.A. | **1080** [77] |
| $Ge_2H_2$ (h) | N.A. | **995** [77] |
| $Ge_2Te_2$ (h) | N.A. | **234** [77] |
| GeTe (h) | N.A. | **172** [77] |
| $Hf_2CO_2$ (h) | N.A. | **190** [77] |
| $HfBr_2$ (h) | N.A. | **588** [77] |
| HfBrCl (h) | N.A. | **240** [77] |
| HfBrI (h) | N.A. | **805** [77] |
| $HfCl_2$ (h) | N.A. | **277** [77] |
| HfClI (h) | N.A. | **482** [77] |
| $HfI_2$ (h) | N.A. | **4782** [77] |
| InAs (h) | N.A. | **372** [77] |
| $Ir_2Br_2O_2$ (h) | N.A. | **309** [77] |
| $Ir_2Br_2S_2$ (h) | N.A. | **488** [77] |
| $Ir_2Cl_2O_2$ (h) | N.A. | **250** [77] |
| $Ir_2Cl_2S_2$ (h) | N.A. | **401** [77] |
| $Ir_2I_2S_2$ (h) | N.A. | **578** [77] |
| $Mo_2Cl_6$ (h) | N.A. | **78** [77] |
| $Pb_2Te_2$ (h) | N.A. | **161** [77] |
| $PtTe_2$ (h) | N.A. | **337** [77] |
| $S_2Si$ (h) | N.A. | **31** [77] |
| $Sb_2$ (h) | N.A. | **2044** [77], 1330[15] |
| $Se_2Si$ (h) | N.A. | **55** [77] |
| $Si2Cl_2$ (h) | N.A. | **406** [77] |
| $Si_2H_2$ (h) | N.A. | 101[20] |
| $Sn_2H_2$ (h) | N.A. | **2063** [77] |
| $Sn_2Te_2$ (h) | N.A. | **238** [77] |
| $Ti_2Br_2N_2$ (h) | N.A. | **156** [77] |
| $Ti_2CO_2$ (h) | N.A. | **347** [77] |
| $TiBr_2$ (h) | N.A. | **753** [77] |
| TiBrI (h) | N.A. | **451** [77] |
| $TiI_2$ (h) | N.A. | **884** [77] |
| WSSe (h) | N.A. | **515** [77] |
| WSeTe (h) | N.A. | **211** [77] |
| $WTe_2$ (h) | N.A. | **349** [77] |
| $Zr_2CO_2$ (h) | N.A. | **118** [77] |
| $ZrBr_2$ (h) | N.A. | **766** [77] |
| ZrBrCl (h) | N.A. | **476** [77] |
| ZrBrI (h) | N.A. | **904** [77] |
| $ZrCl_2$ (h) | N.A. | **370** [77] |
| ZrClI (h) | N.A. | **337** [77] |
| $ZrI_2$ (h) | N.A. | **5138** [77] |